\documentclass[twocolumn]{aastex62}

\usepackage[flushleft]{threeparttable}
\usepackage{apjfonts}
\usepackage{subfigure}
\usepackage{xcolor}
\usepackage{graphicx}

\shorttitle{Transmission Spectrum of Kepler-79d}
\shortauthors{Y. Chachan et al.}

\begin{document}

\title{\textbf{A Featureless Infrared Transmission Spectrum for the Super-Puff Planet Kepler-79d}}
\correspondingauthor{Yayaati Chachan}
\email{ychachan@caltech.edu}

\author[0000-0003-1728-8269]{Yayaati Chachan}
\affil{Division of Geological and Planetary Sciences, California Institute of Technology, 1200 E California Blvd, Pasadena, CA, 91125, USA}

\author[0000-0002-6227-7510]{Daniel Jontof-Hutter}
\affil{Department of Physics, University of the Pacific, 3601 Pacific Avenue, Stockton, CA 95211, USA}

\author{Heather A. Knutson}
\affil{Division of Geological and Planetary Sciences, California Institute of Technology, 1200 E California Blvd, Pasadena, CA, 91125, USA}

\author[0000-0001-9897-9680]{Danica Adams}
\affil{Division of Geological and Planetary Sciences, California Institute of Technology, 1200 E California Blvd, Pasadena, CA, 91125, USA}

\author[0000-0002-8518-9601]{Peter Gao}\altaffiliation{51 Pegasi b Fellow}
\affiliation{Department of Astronomy, University of California, Berkeley, CA 94720, USA}

\author[0000-0001-5578-1498]{Bj\"orn Benneke}
\affiliation{Department of Physics and Institute for Research on Exoplanets, Université de Montréal, Montreal, QC, Canada}

\author{Zachory Berta-Thompson}
\affiliation{Department of Astrophysical and Planetary Sciences, University of Colorado at Boulder, Boulder, CO 80309, USA}

\author[0000-0002-8958-0683]{Fei Dai}
\affil{Division of Geological and Planetary Sciences, California Institute of Technology, 1200 E California Blvd, Pasadena, CA, 91125, USA}

\author{Drake Deming}
\affiliation{Department of Astronomy, University of Maryland at College Park, College Park, MD 20742, USA} 
\affiliation{NASA Astrobiology Institute's Virtual Planetary Laboratory, USA}

\author[0000-0001-6545-639X]{Eric B. Ford}
\affiliation{Department of Astronomy and Astrophysics, 525 Davey Laboratory, The Pennsylvania State University, University Park, PA, USA}
\affiliation{Center for Exoplanets \& Habitable Worlds, 525 Davey Laboratory, The Pennsylvania State University, University Park, PA, USA}
\affiliation{Institute for Computational \& Data Sciences, The Pennsylvania State University, University Park, PA, USA}

\author[0000-0002-1228-9820]{Eve J. Lee}
\affiliation{Department of Physics and McGill Space Institute, McGill University, 3550 rue University, Montreal, QC, H3A 2T8, Canada}

\author[0000-0002-2990-7613]{Jessica E. Libby-Roberts}
\affiliation{Department of Astrophysical and Planetary Sciences, University of Colorado at Boulder, Boulder, CO 80309, USA}

\author[0000-0002-4869-000X]{Nikku Madhusudhan}
\affiliation{Institute of Astronomy, University of Cambridge, Cambridge, CB3 0HA, UK}

\author[0000-0003-4328-3867]{Hannah~R. Wakeford}
\affiliation{School of Physics, University of Bristol, HH Wills Physics Laboratory, Tyndall Avenue, Bristol BS8 1TL, UK}

\author[0000-0001-9665-8429]{Ian Wong}\altaffiliation{51 Pegasi b Fellow}
\affiliation{Department of Earth, Atmospheric, and Planetary Sciences, Massachusetts Institute of Technology, Cambridge, MA 02139, USA}

\begin{abstract}
Extremely low density planets (`super-puffs') are a small but intriguing subset of the transiting planet population. With masses in the super-Earth range ($1-10$ M$_{\oplus}$) and radii akin to those of giant planets ($>4$ R$_{\oplus}$), their large envelopes may have been accreted beyond the water snow line and many appear to be susceptible to catastrophic mass loss. Both the presence of water and the importance of mass loss can be explored using transmission spectroscopy. Here, we present new \emph{HST} WFC3 spectroscopy and updated \emph{Kepler} transit depth measurements for the super-puff Kepler-79d. We do not detect any molecular absorption features in the $1.1-1.7$ $\mu$m WFC3 bandpass and the combination of \emph{Kepler} and WFC3 data are consistent with a flat line model, indicating the presence of aerosols in the atmosphere. We compare the shape of Kepler-79d's transmission spectrum to predictions from a microphysical haze model that incorporates an outward particle flux due to ongoing mass loss. We find that photochemical hazes offer an attractive explanation for the observed properties of super-puffs like Kepler-79d, as they simultaneously render the near-infrared spectrum featureless and reduce the inferred envelope mass loss rate by moving the measured radius (optical depth unity surface during transit) to lower pressures. We revisit the broader question of mass loss rates for super-puffs and find that the age estimates and mass loss rates for the majority of super-puffs can be reconciled if hazes move the photosphere from the typically assumed pressure of $\sim 10$ mbar to $\sim 10 \; \mu$bar.
\end{abstract}

\section{Introduction}

The \emph{Kepler} telescope was the first observatory with both the sensitivity and temporal baseline to detect small transiting planets at Earth-like distances around Sun-like stars. Amongst the most valuable contributions of the telescope is the discovery of dynamically interacting multi-planet systems spanning a broad range of orbital periods. \emph{Kepler}'s long 4-year baseline allowed us to observe multiple transits of such systems, to record the variations in the planets' orbital period (Transit Timing Variations a.k.a TTVs), and to obtain dynamical mass measurements for planets that were otherwise inaccessible to the radial velocity (RV) technique due to the host stars' dimness. This technique also led to the discovery of an intriguing new class of extremely low density planets (dubbed `super-puffs') that have super-Earth like masses ($\lesssim 10 \mathrm{M}_{\oplus}$) and gas-giant like radii ($\gtrsim 5 \mathrm{R}_{\oplus}$; \citealp{Masuda2014, Jontof-Hutter2014, Ofir2014, Xie2014, Mills2016, Orosz2019, Vissapragada2020}). Their low implied bulk densities (typically $\sim$ 0.1 g/cc) require the possession of a hydrogen-helium envelope that is tens of percent by mass, quite unlike the typical $\sim 1$\% that most super-Earths are inferred to possess \citep{Lopez2014}. This makes super-puffs particularly interesting from a planet formation perspective, as it is unclear how they were able to acquire such large H/He envelopes. \cite{Lee2016} were the first to point out that super-Earth cores could only accrete such large gas envelopes if the gas had a relatively low opacity (i.e., was effectively dust-free) and the planet was located in a cool, low-density region of the disk. Protoplanetary disk models indicate that these conditions were not likely met at the present-day locations of these super-puffs \citep[e.g.][]{Chiang2013, Ikoma2012, Inamdar2015} and so it is hypothesized that these planets could have formed at a more distant location and then migrated inward. If dust opacity somehow becomes negligible, disk conditions (temperature and hence opacity) beyond the ice line could be favorable for formation of super-puffs, possibly enriching them in water relative to super-Earths that formed in situ.

The relatively low densities of super-puffs also make them highly vulnerable to atmospheric mass loss, either due to photoevaporation or Parker wind-like outflow \citep{Lopez2014, Owen2016, Cubillos2017, Wang2019, Gao2020}. The latter mechanism is important for super-puffs because their low gravities result in non-negligible densities at the Bondi radius. This is especially true if the atmospheric pressure corresponding to the observed transit radius (slant optical depth $\tau \sim 1$ surface) is equal to tens or hundreds of mbar, similar to the values inferred for other exoplanets via transmission spectroscopy \citep[e.g.][]{Sing2016}. In this scenario, the implied mass loss rates for some super-puffs should already have caused them to lose their entire envelope. The fact that super-puffs have managed to retain their large envelopes over billions of years suggests that our knowledge of mass loss processes in these atmospheres is incomplete \citep{Owen2016, Cubillos2017, Fossati2017, Wang2019}.

Transmission spectroscopy is a powerful tool that can provide us with new insights into both the compositions of super-puff atmospheres and their corresponding mass loss rates. Super-puffs are favorable targets for transmission spectroscopy: they have relatively low gravity and their low bulk densities suggest that they are unlikely to have atmospheric metallicities higher than a few $100\times$ solar \citep[e.g.][]{Lopez2014, Thorngren2019}; as such, their atmospheric scale heights are comparable to or greater than those of hot-Jupiters despite their relatively cool equilibrium temperatures ($\sim 500$ K). However, the first two super-puffs observed by the \emph{Hubble Space Telescope} (\emph{HST}) appear to have featureless $1.1 - 1.7 \mu m$ transmission spectra \citep{Libby-Roberts2020}. Although \cite{Libby-Roberts2020} could not entirely rule out atmospheric metallicities above 300$\times$ solar for Kepler-51b and d, they argue that high-altitude aerosols provides a more plausible explanation.

In principle these aerosols could be either condensate clouds or photochemical hazes, but the temperature-pressure profiles for most super-puffs are not expected to cross condensation curves in the upper region ($P < 1$ bar) of the atmosphere \citep[e.g.][]{Morley2015, Crossfield2017, Gao2020}. On the other hand, the relatively low ($\sim500$ K) temperatures of these hydrogen-rich atmospheres make them favorable sites for photochemical haze production, which occurs at relatively low pressures (1-10 $\mu$bar) \citep[e.g.,][]{Horst2018, He2018a, He2018, Kawashima2018, Kawashima2019, Adams2019}. Aerosols entrained in an outflowing atmospheric wind could be carried to even lower pressures ($\lesssim 1$ $\mu$bar), significantly reducing the gas density at the $\tau \sim 1$ surface and thus the Bondi radius, leading to a reduction in the mass loss rate \citep{Wang2019, Gao2020}. This offers an explanation for how these planets have managed to retain their hydrogen-rich envelopes to the present day.

Alternative theories that attempt to explain the large radii and correspondingly low densities of super-puffs have also been proposed. \cite{Pu2017} and \cite{Millholland2019} argue that larger internal heat fluxes, due to Ohmic dissipation and obliquity tides respectively, could inflate planetary radii to produce super-puffs. This would reduce the amount of hydrogen-helium required to match the planet's mass and radius. However, it is unclear if this reduction is sufficient to make the super-puffs' hydrogen-helium repository more commensurate with the wider sub-Neptune population. These models also do not satisfactorily resolve the tension between mass loss rates, atmospheric lifetimes, and planetary ages; regardless of the inflation mechanism, puffy planets are still vulnerable to rapid atmospheric mass loss. Although the \cite{Millholland2019} models (based on \citealp{Chen2016} models) include photoevaporative mass loss, they do not include Parker wind mass loss, which tends to be more important for the most vulnerable super-puffs. Moreover, Ohmic dissipation is unlikely to be as important at equilibrium temperatures of $\sim 500-700$ K that are typical for super-puffs \citep{Pu2017}. It has also been suggested that super-puffs may not be puffy planets at all but planets with face-on rings \citep{Piro2020}. However, this idea has trouble providing a unifying explanation for all super-puffs and is difficult to verify observationally. In this work, we assume that super-puffs do possess large hydrogen-helium envelopes and will comment on these alternative explanations when the need arises.

In this paper we examine the super-puff Kepler-79d, a planet on a 52 day orbit around an F-type star \citep{Jontof-Hutter2014}. Kepler-79 has four dynamically interacting planets with periods that are near a 1:2:4:6 chain of commensurability, which allows us to derive planet masses from transit timing variations. All planets in this system have masses in the super-Earth regime ($\lesssim 10 \mathrm{M}_{\oplus}$) and relatively large radii (varying from $3.5 - 7 \mathrm{R}_{\oplus}$), implying low bulk densities and a significant volatile envelope. In particular, Kepler-79d has a mass of 5.3 M$_{\oplus}$ and a radius of 7 R$_{\oplus}$, with corresponding bulk density of 0.08 g/cc, placing it firmly in the super-puff regime. Kepler-79, with an estimated age of $1.3^{+1.0}_{-0.4}$ Gyrs \citep{Fulton2018}, is most likely older than Kepler-51 ($0.5 \pm 0.25$ Gyrs, \citealp{Libby-Roberts2020}). As a result, we expect that the planets in the Kepler-79 system are less likely to be appreciably inflated by residual heat from their formation than those in the Kepler-51 system. This is because most of the contraction happens in the first few 100 Myrs \citep[e.g.,][]{Lopez2014, Libby-Roberts2020}. This means that Kepler-79d's anomalously large radius and low density can only be matched with a high gas-to-core mass fraction \citep[$\sim$36\%,][]{Lopez2014}. Kepler-79 also appears to be less active than Kepler-51, with a low variability amplitude in the \emph{Kepler} bandpass ($<0.2\%$ compared to $\sim 1.2\%$ for Kepler-51; \citealp{McQuillan2014}; see \S~\ref{sec:stellar_activity}) and no evidence for spot crossings in the \emph{Kepler} transit light curves of Kepler-79d (as opposed to Kepler-51b and d: $17\%$ of their \emph{Kepler} transits show spot crossings by eye; \citealp{Libby-Roberts2020}). This makes it less likely (relative to Kepler-51) that the planet's transmission spectrum will be significantly affected by stellar activity.

Here, we present new \emph{HST} WFC3 transit spectroscopy for Kepler-79d spanning the $1.1-1.7$ $\mu$m wavelength range and combine our analysis with previously published \emph{Kepler} data in a self-consistent framework. We describe our data reduction and light curve fitting routines in \S\ref{sec:obs} and \ref{sec:lc_fitting}. The resulting white light curve depths and updated mass estimates from a transit timing variation (TTV) analysis are presented in \S\ref{sec:wlc} and \S\ref{sec:ttv}. In \S\ref{sec:transmission_spectrum}, we use the shape of the observed transmission spectrum to place constraints on Kepler-79d's atmospheric composition and aerosol properties. We also present models for Kepler-79d generated using a modified version of the Community Aerosol and Radiation Model for Atmospheres (CARMA) to study haze formation and entrainment in the outflowing atmospheric wind. In \S\ref{sec:puffy_mass_loss}, we examine the mass loss rates for the super-puff population as a whole and discuss the implications in light of the host stars' ages. Finally, in \S\ref{sec:future_conclusions} we present our conclusions and discuss potential future observations.

\section{Observational Data}     \label{sec:obs}
\subsection{\emph{HST} WFC3 Observations and Spectral Extraction}
We observed transits of Kepler-79d with \textit{HST}'s Wide Field Camera 3 (WFC3) instrument on UT 2018 April 12 and UT 2018 November 6 (PI Jontof-Hutter, GO 15138). This relatively long period (52 days) planet has an approximately eight hour transit duration, and each visit therefore consisted of 13 \emph{HST} orbits in order to ensure that that our out-of-transit baseline was comparable to the time in transit. The long duration of these observations meant that \emph{HST} inevitably crossed the South Atlantic Anomaly (SAA) during a few orbits in each visit, however its impact on our data appears to be minimal as we discuss below.

\begin{figure}
    \centering
    \includegraphics[width=\linewidth]{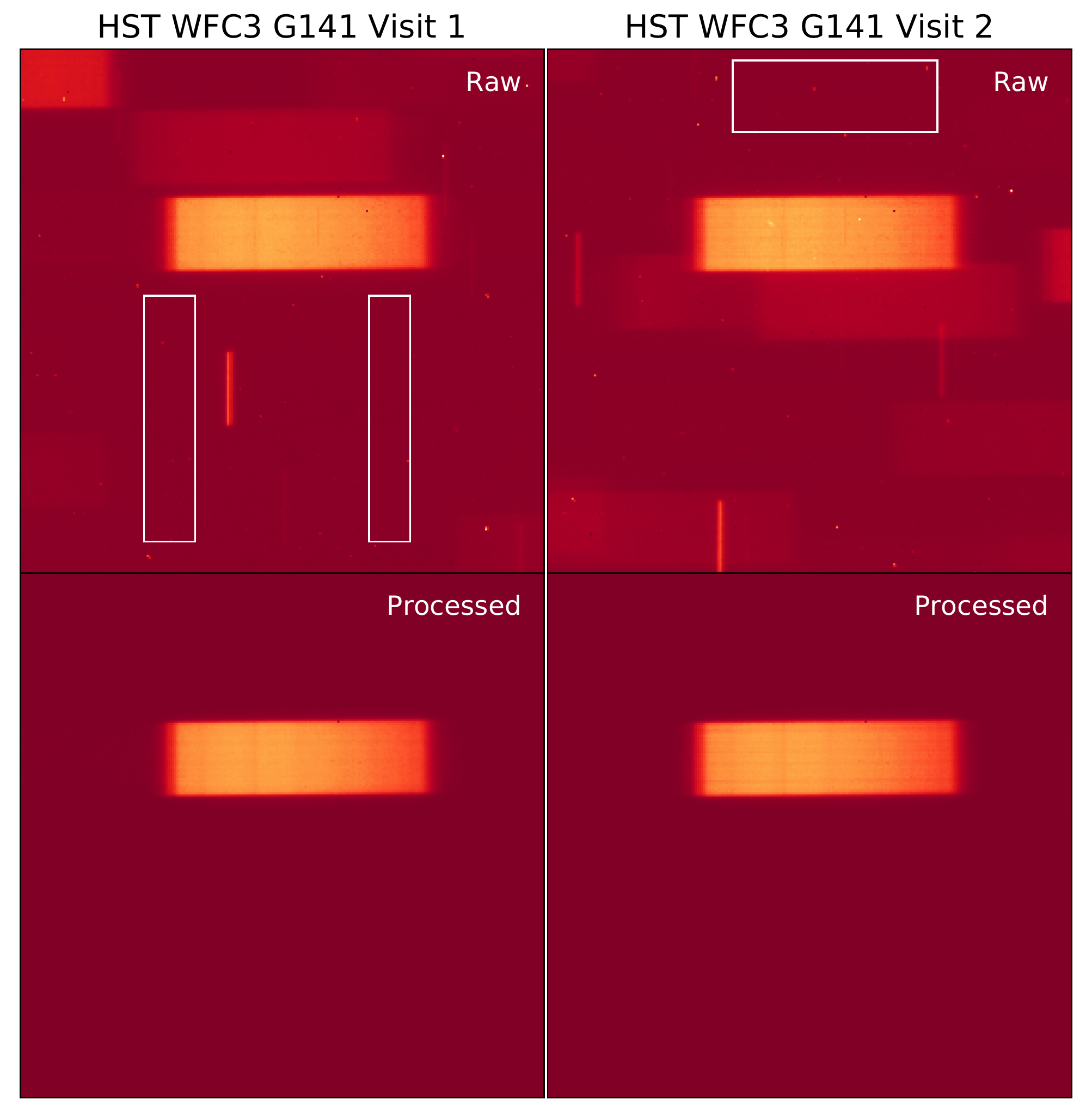}
    \caption{First exposures of the 2 \emph{HST} visits. The top panel contains the raw images and the bottom panel contains the processed images (after extraction and outlier correction. The region used for sky background calculation is marked out by a white box in the raw images. Kepler-79 has the brightest spectral trace in these images.}
    \label{fig:raw_images}
\end{figure}

The data were taken with the G141 grism in the 256$\times$256 sub-array mode. We utilized the unidirectional (forward scan only) spatial scan mode in order to increase the duty cycle for these observations relative to the more conventional staring mode \citep{Mccullough2012, Deming2013}. Although observations of the brightest ($J\lesssim10$) stars typically alternate forward and reverse scans \citep[e.g.,][]{Knutson2014a, Stevenson2014, Mansfield2018}, this would have required us to fit two independent instrumental noise models, one for each scan direction. For Kepler-79 ($J=12.9$), the difference in duty cycle for forward-only versus forward and reverse scans was negligible, and we therefore opted for the simpler unidirectional scan mode. The orientation of the spacecraft during data collection and the scan length ($4.46''$ with a scan rate of $0''.015$ s$^{-1}$) were set to ensure that the spectrum of Kepler-79 did not overlap with those of any neighbouring stars (Figure~\ref{fig:raw_images}). For this relatively faint star each exposure lasted 290.8 seconds, yielding 7 exposures per \emph{HST} orbit. Kepler-79 is only visible for approximately half of \emph{HST}'s 96 minute orbit, leading to gaps in our observations. During the first visit, one orbit covered part of ingress, but none of the orbits in the second visit covered ingress or egress. As a result, the transit time for the first visit from our white-light fits is much better constrained than the transit time for the second visit.


We use the \texttt{ExoTEP} suite for our data reduction, which is described in detail in \cite{Benneke2019} and follows the methods previously adopted in \cite{Berta2012}, \cite{Deming2013}, and \cite{Knutson2014c} for WFC3 data. We use bias- and dark-corrected \texttt{ima} images produced by the standard \textit{calwfc3} pipeline. Each exposure consists of 14 non-destructive reads and we subtract consecutive reads to create difference sub-exposures \citep[e.g.][]{Deming2013, Kreidberg2014, Evans2016}. The rows where the median flux profile falls to 20\% of the peak flux value in the cross-dispersion direction delineate the vertical extent of the sub-exposure. We find that including the flux from an additional buffer of pixels above and below these rows reduces the correlated noise in the white light curves. We optimize this buffer by picking the values (9 pixels and 10 pixels for the first and second visit respectively) that minimize the scatter in the residuals in our light curve fits and ensure by visual inspection that secondary sources are not included.

We estimate the sky background using two $120 \times 20$ pixel boxes below the spectral trace in the first visit and one $35 \times 100$ pixel box above the trace in the second visit (Figure~\ref{fig:raw_images}). We ensure that these regions do not contain secondary sources and remove $3\sigma$ outliers before subtracting the median of the remaining pixels from the sub-exposure. These background subtracted sub-exposures are then combined to form full frame images. We flat-field all frames using the calibration files provided by STScI \citep{Kuntschner2011} following the method outlined in \cite{Wilkins2014}. We quantify image-to-image variations in the position of the spectral trace in the $x$ (dispersion) direction by by summing each image in the $y$ direction and using the first summed image as a 1D spectral template to calculate the $x$ offset of all the subsequent images. The 2D wavelength solution is then calculated for each image with the method outlined in \cite{Tsiaras2016} and \cite{Benneke2019} using the wavelength and trace calibration functions provided by STScI \citep{Kuntschner2009}.

Because Kepler-79 is fainter and our exposure times are longer than in previous studies utilizing the \texttt{ExoTEP} pipeline \citep{Benneke2019, Chachan2019, Benneke2019a, Wong2020}, we find that we require a more robust outlier recognition and replacement method to correct for cosmic rays and bad pixels. We do a first pass filtering step to flag obvious outliers using the same spatial outlier correction used in previous studies.  In this case, we make two passes with a moving median filter (11 pixels by 11 pixels) where we flag 6$\sigma$ outliers and replace them by the median value in each image. Although we experimented with lower $\sigma$ thresholds, we found that they led to overly aggressive spatial outlier correction.

We identify and correct any remaining outliers using the spatio-temporal filtering method outlined in \cite{Nikolov2014} and \cite{Nikolov2018}. In this step, we subtract the two preceding and two succeeding images from the current exposure to construct four difference images. We then take the median of these four difference images and flag 5$\sigma$ outliers in this median difference image using the same 2D moving median filter as before. For each outlier, we then construct a median `PSF' profile in the cross-dispersion (scan) direction from the five preceding and succeeding columns in the image. This median `PSF' profile is scaled to match the median flux level in the column with the outlier. The outlier is then replaced with the corresponding flux value at that pixel location in the scaled median `PSF' profile.  This method results in more accurate replacement flux values than a simple spatial median because it is better able to account for variations in the scan rate of the telescope as  it moves across the detector. We find that two iterations with this filter are enough to remove visible outliers from all of our exposures.

We use the 2D wavelength solution to determine the boundaries of 30 nm wide bins and sum the flux from both fully and partially included pixels to obtain a 1D spectrum. For the partial pixels, we use a flux-conserving second order 2D polynomial to calculate the contribution of flux to that particular bin (see \citealp{Tsiaras2016} for more details). The white light curve is obtained by summing the flux from all the spectroscopic light curves ($1.12 - 1.66 \mu$m).

\subsection{Kepler Light Curves}
\label{sec:kepler_lc}
The \emph{Kepler Space Telescope} observed 28 transits of Kepler-79d between 2009 and 2013. Most of these data were obtained in short cadence (1 min integrations) mode, but during the first, second, and seventeenth quarters, only long cadence (30 min integrations) data were collected. In this study, we utilize the short cadence simple aperture photometry (SAP) light curves (24 transits in total), as these data provide better information about the transit shape than the long cadence observations. The 10th, 16th, and 17th transit data contain significant correlated noise with an estimated magnitude larger than 100 ppm (see \S~\ref{sec:wlc}) and we therefore exclude them when we create our phased \emph{Kepler} light curve. Since the ephemeris for Kepler-79d is not linear, we utilize individual mid-transit times from \cite{Jontof-Hutter2014} to extract sections of the light curve centered around each transit event with a length of three times the transit duration. We then fit for a quadratic trend in the out of transit baseline and remove it from each transit before combining all the transits to form a single light curve centered on the mid-transit phase.

We find that linear detrending is inadequate to fit the out-of-transit baseline; this is unsurprising given the relatively long duration ($\sim 24$ hours) of our extraction window. When we compare quadratic and linear detrending, we find that quadratic detrending is highly favored by the Bayesian information criterion ($\Delta$ BIC = 74) and reduces the scatter in the residuals by $1 \sigma$. After creating our phased transit light curve, we perform outlier correction on it by using a moving median filter in two steps. First, we perform three iterations of outlier rejection using a moving median filter with a width of 20 exposures and a relatively high 5$\sigma$ threshold. We then repeat this outlier rejection using a filter with a 50 exposure width and trim any points that deviate from this moving median by more than $3 \sigma$. This second step flags just 0.06\% of all the short cadence data points, and therefore has a negligible effect on the best-fit transit shape.

\section{Light Curve Modelling and Fitting}
\label{sec:lc_fitting}
\subsection{Astrophysical Model}
We use the \texttt{BATMAN} package \citep{Kreidberg2015} to model transit light curves and fit for the planet-star radius ratio $R_p/R_*$, mid-transit time $T_c$, impact parameter $b$, and semi-major axis to stellar radius ratio $a/R_*$. We calculate custom stellar limb darkening coefficients for the \emph{HST} WFC3 bandpass using the package \texttt{LDTk} \citep{Parviainen2015}, which uses the \texttt{PHOENIX} stellar spectra models \citep{Husser2013}. \texttt{LDTk} generates radial stellar brightness profiles and then fits these profiles with a fourth order non-linear limb darkening model. The stellar properties are taken from \cite{Petigura2017a} and \cite{Fulton2018}, and are derived using \emph{Gaia} parallaxes, \emph{Kepler} photometry, and spectroscopic temperatures from Keck/HIRES. For Kepler-79, this study finds $T_{\mathrm{eff}} = 6389 \pm 60$ K, [Fe/H] = $0.06 \pm 0.04$, and log $g = 4.33 \pm 0.10$. Since the \emph{Kepler} light curve contains dense sampling of the transit shape, we fit for quadratic limb darkening coefficients instead of fixing them to the model values from \texttt{LDTk} \citep[also recommended in the literature, e.g.][]{Espinoza2015}. \texttt{ExoTEP} allows for a quadratic and a four parameter limb darkening law and we verified that using the latter does not improve the fit. Our fitted quadratic limb darkening coefficients (listed in Table~\ref{table:wlc_fits}) are consistent within $1 \sigma$ with those obtained from \texttt{ATLAS} models in J-band \citep{Kurucz1979, Claret2011}. Although the two limb darkening coefficients obtained from \texttt{LTDk} (which uses \texttt{PHOENIX} models that are more suitable to cooler stars) are $3 \sigma$ and $1 \sigma$ off from our fitted values, this does not introduce any wavelength dependent bias in our analysis as the difference in the limb darkening between \texttt{ATLAS} and \texttt{PHOENIX} models in the \emph{HST} WFC3's infrared bandpass is negligible compared to the uncertainties in the measured flux.

\begin{table*}
	\begin{threeparttable}
    	\caption{Global Broadband Light Curve Fit Results} \label{table:wlc_fits}
    	\begin{center}
        	\begin{tabular*}{\textwidth}{@{\extracolsep{\fill}} c c c c}
        		\hline \hline
        		Parameter & Instrument & Band pass ($\mu$m) & Value \\ \hline 
        		Planet radius, $R_p/R_*$ & \emph{Kepler} & 0.42 - 0.9 & $0.04979_{-0.00021}^{+0.00027}$     \\
        		Transit depth, $(R_p/R_*)^2$ (ppm) & \emph{Kepler} & 0.42 - 0.9 & $2478.9_{-20.9}^{+27.0}$     \\
        		Planet radius, $R_p/R_*$ & WFC3 G141 & 1.1 - 1.7 &   $0.04876 \pm 0.00046$       \\
        		Transit depth, $(R_p/R_*)^2$ (ppm) & WFC3 G141 & 1.1 - 1.7 &   $2377.7_{-44.3}^{+45.0}$       \\
        		Transit center time $T_c$ (BJD$_{\mathrm{TDB}}$)\textsuperscript{a} & WFC3 G141 (Visit 1) & 1.1 - 1.7 & $2458221.38634_{-0.00092}^{+0.00093}$ \\
        		Transit center time $T_c$ (BJD$_{\mathrm{TDB}}$)\textsuperscript{a} & WFC3 G141 (Visit 2) & 1.1 - 1.7 & $2458429.7253_{-0.0066}^{+0.0074}$ \\
        		Impact parameter $b$ & -- & -- & $0.16_{-0.11}^{+0.12}$ \\
        		Relative semi-major axis $a/R_*$ & -- & -- & $47.04_{-1.23}^{+0.51}$ \\
        		Inclination\textsuperscript{b} $i$ & -- & -- & $ 89.81_{-0.16}^{+0.13}$ \\
        		Limb darkening coefficient $u_1$ & \emph{Kepler} & 0.42-0.9 & $0.25 \pm 0.06$ \\
        		Limb darkening coefficient $u_2$ & \emph{Kepler} & 0.42-0.9 & $0.33 \pm 0.11$ \\
        		\hline
        	\end{tabular*}
        	 \begin{tablenotes}
             \small
			 \item {\bf Notes.} 
			 \item \textsuperscript{a}{Subtract 69.184 seconds to convert to BJD$_{\mathrm{UTC}}$ \citep[see][]{Eastman2010}.}
			 \item \textsuperscript{b}{Calculated from posteriors for $b$ and $a/R_*$.}
			\end{tablenotes}
    	\end{center}
	\end{threeparttable}
\end{table*}

\subsection{HST/WFC3 Systematics Model}
We fit the white-light curve for each \emph{HST} visit using a linear plus exponential function of the orbital phase ($t_{orb}$) and the $x$ position of the spectral trace on the detector (relative to the first exposure's position $x_o$). We also include an exponential function of time since beginning of visit ($t_v$). These exponential terms are needed in order to correct for charge-trapping in the array \citep[e.g.][]{Deming2013, Zhou2017}. Our WFC3 systematics model $S (t)$ is then: 
\begin{equation}
S (t) = \left( c + p \; t_{orb} + m \; (x - x_o) \right)  \left( 1-e^{-at_{orb}-b-dt_v} \right)
\label{eq:wfc3_systematics}
\end{equation}
where $c$, $p$, $m$, $a$, $b$, and $d$ are free parameters in the fit. The parameters $c$, $p$ and $m$ characterize the linear dependence of systematic noise on $t_v$ and the $x$ position of the spectral trace. For the exponential ramp, $b$ sets the overall time-independent amplitude of the exponential term, and $a$ and $d$ control the dependence on $t_{orb}$ and $t_v$ respectively \citep[e.g.][]{Berta2012, Knutson2014c}. We find that including a visit-long ramp (exponential term in $t_v$) along with the classic orbit-long ramp significantly improves our fit to the systematics in the data ($\Delta$ BIC = 160 and 277 for the first and second visit respectively). In addition, this exponential term in $t_v$ is preferred over the more typically utilized polynomial functions of $t_v$ (e.g. $\Delta$ BIC = 145 and 261 for the first and second visit respectively, for a linear $t_v$ function as opposed to the exponential ramp we use). Although \emph{HST} WFC3 phase curve observations \citep[e.g.,][]{Stevenson2014, Kreidberg2018} often have observational baselines with a length comparable to that of our Kepler-79 observations, they typically observe stars much brighter than Kepler-79. Because the timescale for charge-trapping increases for faint stars, it is unsurprising that the initial exponential ramp would persist across multiple orbits, whereas for bright stars it is typically converged by the end of the first orbit. We also consider fits with an additional linear trend in $t_v$, but found that this does not improve the fit and is disfavoured by BIC.

It is common practice to discard the first exposure in each orbit and the first orbit in each visit in \emph{HST} transit observations, as the very steep rise in flux during these sections of the light curve is typically not well matched by the simple polynomial and/or exponential functions used to approximate trends due to charge-trapping and other spacecraft systematics \citep[e.g.,][]{Deming2013, Sing2016, Tsiaras2018}. For this reason, we also discard the first exposure of each orbit for both of our visits and the first orbit of the first visit. For the second visit, we find that the first orbit is well-matched by our exponential model and its inclusion or exclusion does not bias our estimates of the astrophysical parameters, and we therefore include it in our fits.

For the wavelength-dependent light curves, we consider two different instrumental noise models. The first model involves fitting the full systematics model (Equation~\ref{eq:wfc3_systematics}) to each spectroscopic light curve. In the second model, we apply a common-mode correction to the light curve before fitting a linear function of the $x$ (dispersion direction) position of the each exposure. For the common mode correction, we divide each spectroscopic time series by the ratio of the uncorrected white light curve and the best-fit white light curve transit model \citep[e.g.][]{Deming2013}. The resultant spectroscopic time series is fit with a systematics model that depends on just two parameters, an offset $f$ and a slope $v$ for the detrending parameter $x$:
\begin{equation}
    S(t) = f + v \; (x - x_o)
\end{equation}
We find that the second instrumental noise model is strongly favored by BIC ($\Delta$ BIC in the range $14-73$ for the 18 spectroscopic light curves), and therefore use it for our final analysis of the spectroscopic light curves.

\begin{figure*}
	\centering
	\includegraphics[width=\textwidth]{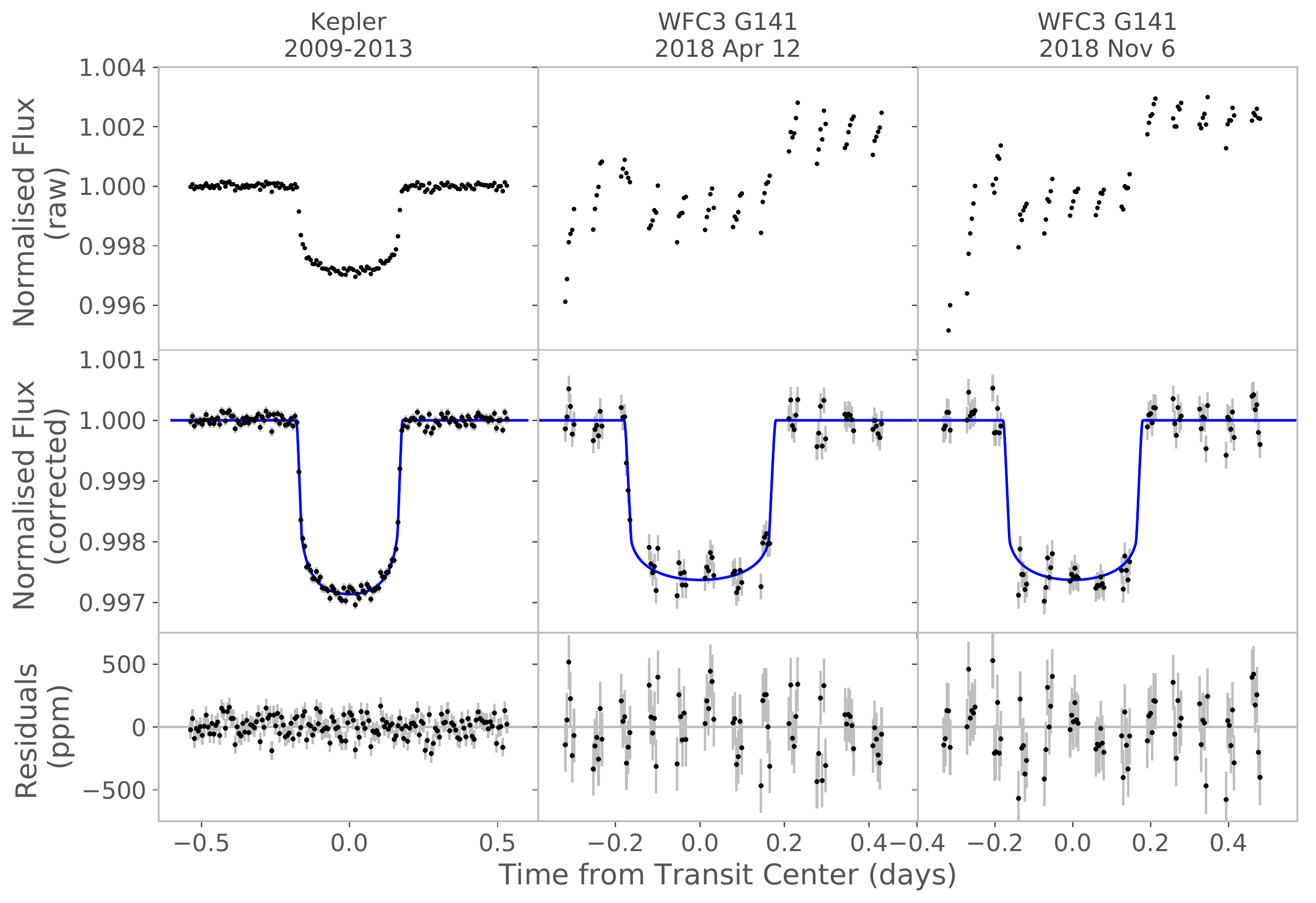}
	\caption{White light transit light curves before (top) and after (middle) dividing out the best-fit instrumental systematics model. The best-fit transit light curve is shown in blue for comparison, and the fit residuals are shown at the bottom. \emph{Kepler} data has been binned down using a bin width of 200 points.}
	\label{fig:WLC_fits}
\end{figure*}

\subsection{Light Curve Fits}
\label{sec:lc_fits}
We initially fit the phased TTV-removed \emph{Kepler} transit light curve (extraction described in \S~\ref{sec:kepler_lc}) and each individual \emph{HST} transit light curve separately and use the best-fit values obtained from these fits as our initial guesses for the joint fit. For the processed \emph{Kepler} light curve, only the astrophysical model is used to fit the data as the systematic trends have already been removed. For the joint fit, we assume that $b$ and $a/R_*$ are the same for all light curves, but allow the mid-transit times for the two \emph{HST} visits to vary independently. Assuming the same $b$ and $a/R_*$ for all visits allows a robust comparison of the transit depths in different bandpasses. We fit for two separate $R_p / R_*$ values, corresponding to the measured transit depths in the \emph{Kepler} and \emph{HST} WFC3 bandpasses. We do not fit for the orbital period, eccentricity, or the argument of periastron in our default fit and instead fix these parameters to the values reported in \cite{Jontof-Hutter2014}. We fit a total of 23 parameters in our joint fit using the affine-invariant ensemble sampler \texttt{emcee} package \citep[v2.2.1, ][]{Foreman-Mackey2012}. The number of walkers is set equal to 4 $\times$ the number of parameters (e.g., 92 walkers for our joint fit). 

For the individual fits of the light curves, we run a 4,000 step burn-in chain followed by a 6,000 step chain that is used to obtain initial guesses for the joint fit of the light curves. Using the burn-in chain, we identify and discard any walkers that get trapped in local minima: if any walker's maximum likelihood is less than the median likelihood of any of the other walkers, we discard it. For the joint fit, we perform two independent fits to the light curves. In the first fit, we run a 40,000 step burn-in chain and an additional 60,000 step chain thereafter to obtain parameter estimates. We then initiate a second fit to the light curves by setting the initial positions of the walkers to within $1 \sigma$ of the best-fit solution from the first fit. For this second fit, we again run a 40,000 step chain for burn-in to ensure that the spread in the walkers' positions equilibrates. After the burn-in, we run a 400,000 step chain to obtain our posteriors and parameter estimates.

To check for convergence in our joint fit, we plot a histogram of likelihoods for individual walkers and find that they all have similar peaks, i.e., all walkers have found the correct global maximum likelihood by the end of burn-in. The parameter estimates we obtain from our two independent joint fits agree at better than the $0.05 \sigma$ level. We also calculate the autocorrelation length ($\xi$) for each walker and variable from our 400,000 step chain using the autocorrelation calculator provided in v3.0.2 of \texttt{emcee}. On average, $\xi$ for a given walker is a factor of $400-500$ times smaller than the chain length for all but three parameters. The strong degeneracy between $a/R_*$ and $b$ leads to longer $\xi$ for these parameters such that the chain length is $\sim 130-150$ times their $\xi$ on average. This degeneracy also lengthens $\xi$ for $R_p/R_*$ in the \emph{Kepler} bandpass, but with a chain length equal to $167 \;  \xi$ on average, our estimate is reliable at the requisite confidence level.

\section{Results and Discussion}
\subsection{White Light Curve Fits}    
\label{sec:wlc}
Results from our global fit to the \emph{Kepler} and \emph{HST} WFC3 white light curves are tabulated in Table~\ref{table:wlc_fits} and the raw and fitted light curves are shown in Figure~\ref{fig:WLC_fits}. Our best-fit orbital parameters and \emph{Kepler} planet-star radius ratio agree with those published in \cite{Jontof-Hutter2014} at the $\sim 1 \sigma$ and$\sim 2 \sigma$  level respectively. Our measured radius ratio in the \emph{Kepler} band is approximately 4\% ($2.2 \sigma$) larger than the corresponding radius ratio in the \emph{HST} WFC3 band.  This allows us to place constraints on the magnitude of potential signatures from scattering (\S~\ref{sec:carma}) and stellar activity (\S~\ref{sec:stellar_activity}).

We test for the presence of time variability in the transit shape by re-fitting individual \emph{Kepler} and \emph{HST} transits with orbital parameters and limb darkening coefficients (for the \emph{Kepler} transits) fixed to the best-fit values from the global fit. Light curves for individual transits in the \emph{Kepler} bandpass are extracted using mid-transit times from \citep{Jontof-Hutter2014} and are subject to the same outlier correction method that is used for the phase folded light curve (see \S~\ref{sec:kepler_lc} for more details). Unlike in the global fit, we do not detrend the \emph{Kepler} data prior to fitting the transit. Instead we simultaneously fit a quadratic function of time along with the transit light curve. We find that this simultaneous baseline and transit fit increases the average uncertainty on individual transit depths by approximately 40\% as compared to fits where we detrend the data first and fit the transit afterward.

We fit all 24 \emph{Kepler} transits using the method described above and find that there is one transit (the 10th) that appears to be significantly deeper than the other transits.  We investigate whether or not this could be due to time-correlated noise in the transit light curve as follows.  First, we estimate the magnitude of the correlated noise in each individual transit light curve by fitting the standard deviation of the residuals ($\sigma$) as a function of bin size ($N$) with a two component model: $\sigma = \sqrt{\sigma_w^2 / N + \sigma_r^2}$. Here, $\sigma_w$ and $\sigma_r$ are the white (Gaussian) and correlated noise components respectively \citep[e.g.,][]{Pont2006}. Our transit light curves have $\sigma \sim 1000$ ppm at 1 min cadence, and we set a threshold of $\sigma_r > 100$ ppm (10\% excess) for flagging transits with significant correlated noise.  When we bin the data on 30 min timescales (comparable to the timescale of ingress or egress), this means that the red noise is a significant fraction ($\geq$ 50\%) of the white noise component. 

\begin{figure}
	\centering
	\includegraphics[width=\linewidth]{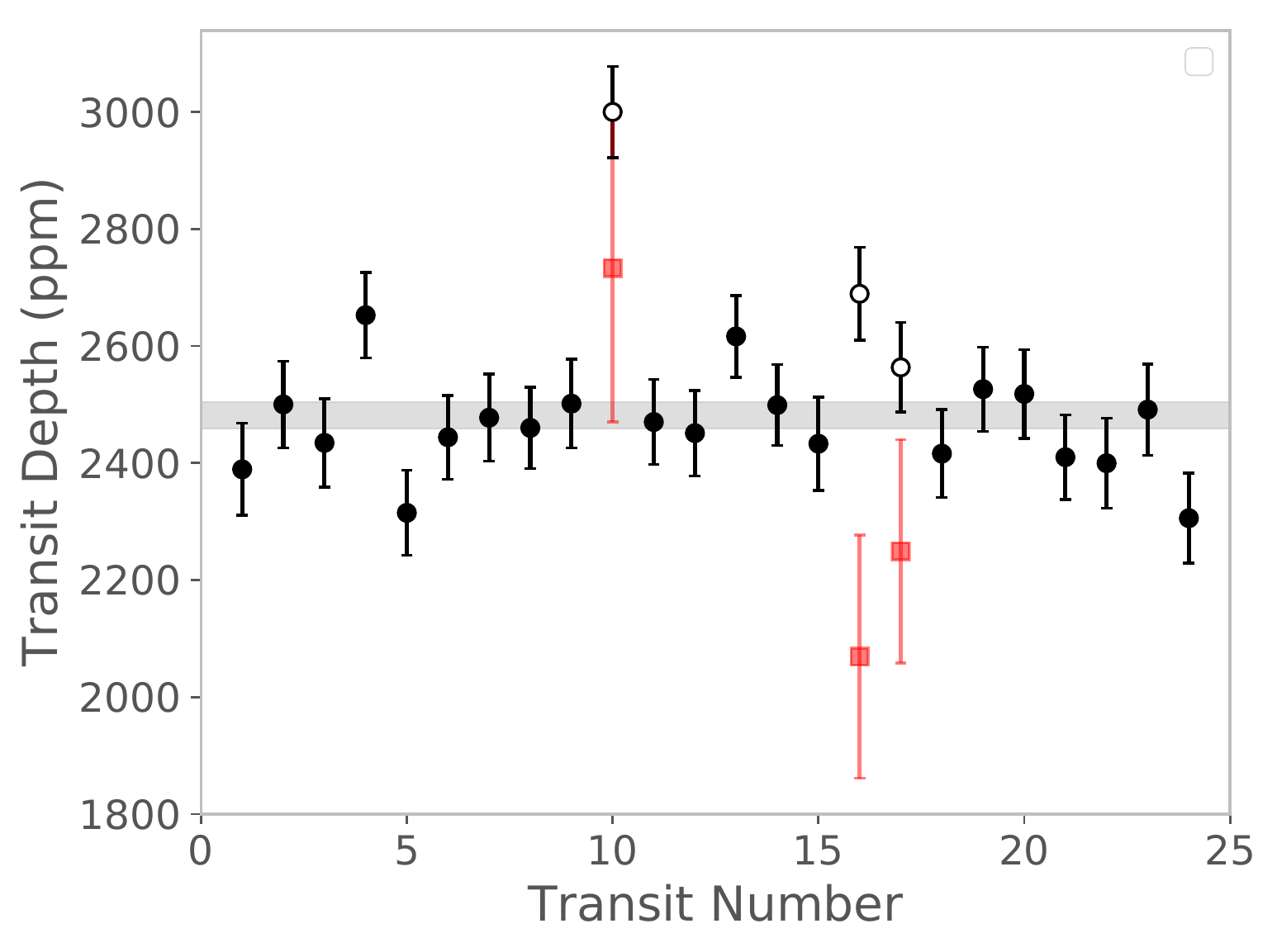}
	\caption{Transit depth measurements for each individual transit of Kepler-79d observed at short cadence. The grey region marks the $1 \sigma$ limits on the transit depth from our joint fit to the \emph{Kepler} and \emph{HST} data. Black empty circles mark the transits that were significantly affected by correlated noise and red squares show their transit depth measurements obtained using Gaussian Process modelling.}
	\label{fig:kepler_depths}
\end{figure}

We find that 10th, 16th, and 17th transits all have red noise levels that exceed this threshold, and we therefore exclude these transits from our phased \emph{Kepler} light curve and variability analysis. The remaining 21 \emph{Kepler} transit depths do not display any significant epoch to epoch variability (reduced $\chi^2$ value of $1.25$, Figure~\ref{fig:kepler_depths}). This stands in contrast to the large epoch to epoch variability observed in the measured \emph{Kepler} transit depths of Kepler-51b and d, which \cite{Libby-Roberts2020} attribute to stellar activity. This lack of variability is in good agreement with the lack of detectable photometric variability for Kepler-79 ($< 0.2\%$) and the absence of any obvious spot crossing events in the \emph{Kepler} light curves.

For the three transits with significant correlated noise, we used Gaussian Processes (GP) modeling to obtain improved estimates of their transit depths and corresponding uncertainties. We fixed the orbital parameters and limb darkening coefficients to the best-fit global values as before and allowed $R_p/R_*$ to vary as a free parameter in the fit. We do not include a quadratic function of time to detrend the data, as the GP is able to fit these trends as part of its noise model. We adopted a squared exponential kernel:
\begin{equation}
    C_{ij} = h^2 \mathrm{exp} \bigg[ - \frac{(t_i - t_j)^2}{2 \eta^2} \bigg] + \sigma^2 \delta_{ij}
\end{equation}
where $C_{ij}$ are elements of the covariance matrix, $t_i$ is the time of the $i$th observation, $\delta_{ij}$ is the Kronecker delta function, $h$ is the amplitude of the covariance, $\eta$ is the correlation timescale, and $\sigma$ is an additional white noise component. We show the resulting transit depth estimates from these fits using red squares in Figure~\ref{fig:kepler_depths}. As expected, the increased uncertainties obtained using GP reflect the presence of significant correlated noise in the data for these three transits. As a check, we also fit two randomly selected transits with low levels of red noise (1 and 5) using GP and confirm that their transit depths are consistent within $1 \sigma$ and their uncertainties increase only by $10-20$\% relative to the values we obtained with a simple quadratic baseline fit. The 21 transits depths combined with the depths of these three transits with significant red noise (obtained using GP) do not display significant variability either (slightly higher reduced $\chi^2$ of $1.35$).

We find that the transit depths for the two \emph{HST} visits (visit 1: $2283 \pm 58$ ppm, visit 2: $2465 \pm 56$ ppm) differ by $2.3 \sigma$. This difference is commensurate with expectations from \emph{HST} white light curves (similar differences observed in previous studies, e.g. \citealp{Wakeford2018, Mansfield2018}), which often suffer from residual time-correlated noise and therefore may have modestly underestimated uncertainties when fit assuming white Gaussian noise. This increase in uncertainties is also corroborated by analyses that use Gaussian Processes instead of parameteric models to fit for the systematic noise in \emph{HST} light curves \citep[e.g.][]{Gibson2012, Evans2018, Evans2019}. We find that the difference in the transit depths between the two \emph{HST} visits is comparable in magnitude to the difference between the averaged \emph{HST} and \emph{Kepler} transit depths, further reinforcing our conclusion  that our data appear to be consistent with a flat line.

\begin{table*}
	\begin{threeparttable}
	\caption{Transit Time Variations Fit Results} \label{table:ttv_fits}
	\begin{center}
    \begin{tabular*}{\textwidth}{@{\extracolsep{\fill}} l c c c c}
      \hline \hline
      Properties & Kepler-79b & Kepler-79c & Kepler-79d & Kepler-79e \\ \hline
      Mass Ratio ($M_{\mathrm{p}} / M_*$) (\emph{Kepler} + \emph{HST}) & $1.84^{+1.00}_{-0.62} \times 10^{-5}$ & $1.09^{+0.32}_{-0.27} \times 10^{-5}$ & $1.30^{+0.24}_{-0.21} \times 10^{-5}$ & $9.39^{+1.77}_{-1.68} \times 10^{-6}$ \\
      Mass/M$_{\oplus}$\textsuperscript{a} (\emph{Kepler} + \emph{HST}) & $7.6^{+3.8}_{-2.6}$ & $4.6^{+1.3}_{-1.1}$ & $5.3^{+0.9}_{-0.9}$ & $3.8^{+0.7}_{-0.6}$ \\
      Mass/M$_{\oplus}$\textsuperscript{a} (\emph{Kepler only}) & $7.6^{+4.2}_{-2.7}$ & $4.6^{+1.4}_{-1.1}$ & $5.3^{+1.0}_{-0.8}$ & $3.9^{+0.7}_{-0.7}$ \\
      Radius/R$_{\oplus}$\textsuperscript{b} & $3.51 \pm 0.10$ & $3.76 \pm 0.11$ & $7.15 \pm 0.20$ & $3.53 \pm 0.16$ \\
      Density\textsuperscript{c} (g/cc) & $0.97^{+0.49}_{-0.33}$ & $0.48^{+ 0.14}_{-0.12}$ & $0.08^{+0.02}_{-0.02}$ & $0.48^{+0.09}_{-0.10}$ \\
      Period (days) & $13.48451^{+0.00010}_{-0.00008}$ & $27.4026^{+0.0004}_{-0.0003}$ & $52.0897^{+0.0006}_{ -0.0005}$ & $81.0665^{+0.0007}_{-0.0006}$ \\
      $e$ sin $\omega$ & $-0.0007^{0.0038}_{-0.0033}$ & $-0.026^{+0.012}_{-0.014}$ & $0.044^{+0.015}_{-0.016}$ & $0.015^{+0.013}_{-0.014}$ \\
      $e$ cos $\omega$ & $-0.022^{+0.007}_{-0.010}$ & $-0.028^{+0.011}_{-0.014}$ & $0.014^{+0.024}_{-0.025}$ & $0.007^{+0.018}_{-0.020}$ \\
      $e\textsuperscript{d}$ & $0.022^{+0.010}_{-0.007}$ & $0.038^{+0.019}_{-0.014}$ & $0.051^{+0.016}_{-0.016}$ & $0.025^{+0.013}_{-0.011}$ \\
      $\omega\textsuperscript{d}$ (degrees) & $182^{+10}_{-9}$ & $223^{+12}_{-12}$ & $71^{+33}_{-25}$ & $59^{+66}_{-57}$ \\
      T$_0$ (BJD - 2,454,900) & $784.3061 \pm 0.0009$ & $806.4771 \pm 0.0014$ & $821.0104 \pm 0.0008$ & $802.1269 \pm 0.0019$ \\
      \hline
    \end{tabular*}
	\begin{tablenotes}
             \small
			 \item {\bf Notes.} 
			 \item \textsuperscript{a}{Planet masses calculated using stellar mass $M_* = 1.244^{+0.027}_{-0.042} M_{\odot}$ from \cite{Fulton2018}. Mass estimates from \emph{Kepler} data alone are shown purely for comparison with estimates obtained from the combination of \emph{Kepler} and \emph{HST} data.}
			 \item \textsuperscript{b}{$R_p/R_*$ values for all planets except Kepler-79d are taken from \cite{Jontof-Hutter2014} and the updated value of $R_* = 1.316^{+0.038}_{-0.037} R_{\odot}$ from \cite{Fulton2018} is used to calculate planetary radii.}
			 \item \textsuperscript{c}{Mass estimates from the combination of \emph{Kepler} and \emph{HST} data are used.}
			 \item \textsuperscript{d}{$e$ and $\omega$ are calculated using posteriors of $e \; \mathrm{sin} \; \omega$ and $e \; \mathrm{cos} \; \omega$.}
	\end{tablenotes}
	\end{center}
	\end{threeparttable}
\end{table*}

\begin{figure*}
	\centering
    \includegraphics[width=0.46\linewidth]{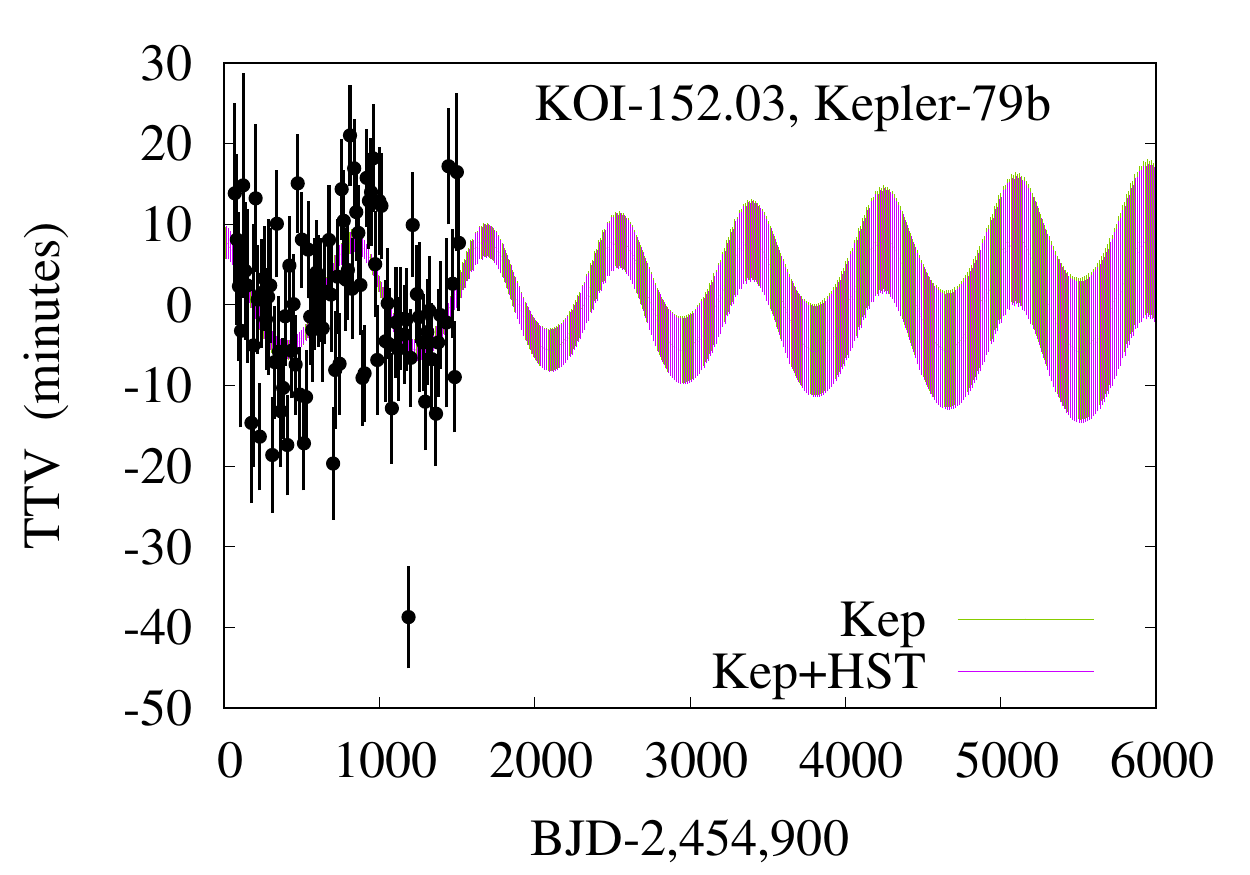}
	\includegraphics[width=0.46\linewidth]{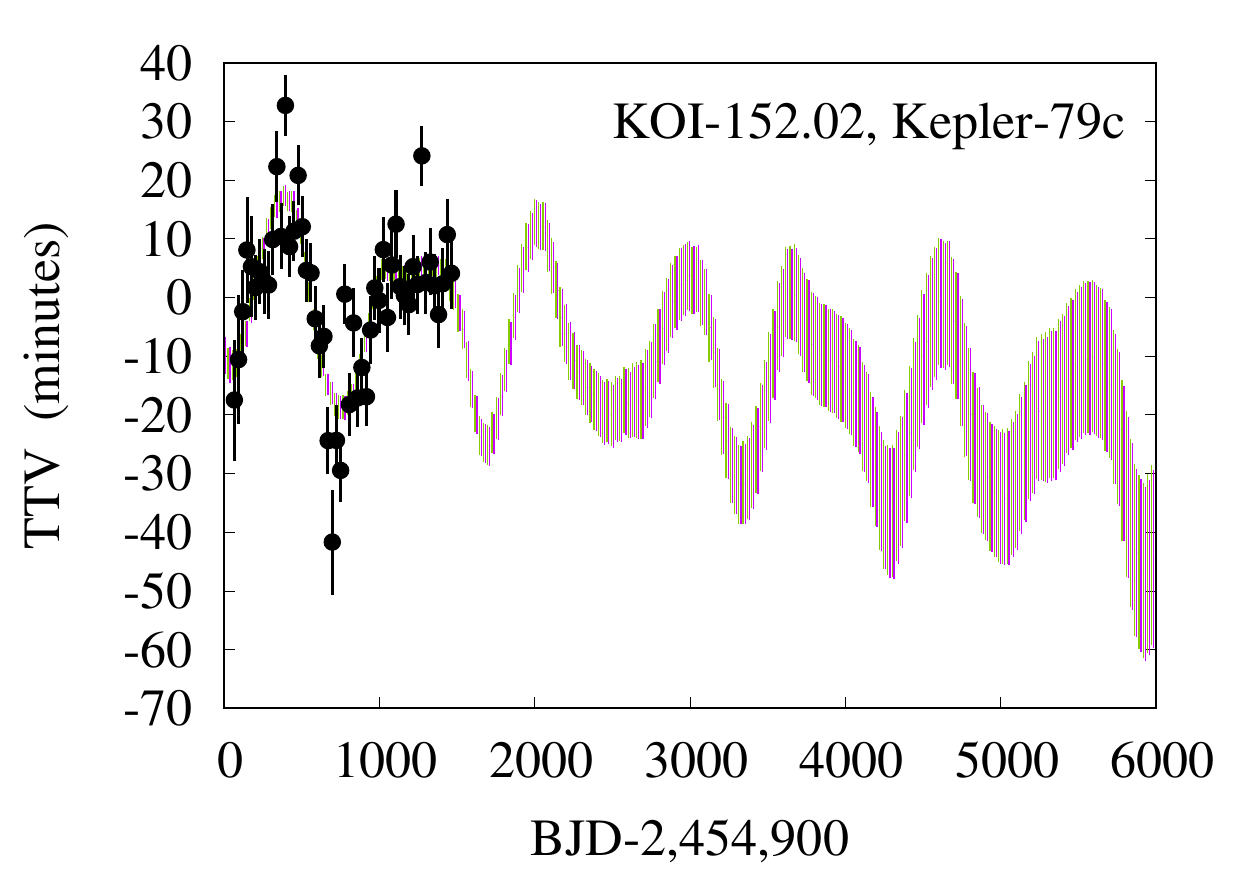}
	\includegraphics[width=0.46\linewidth]{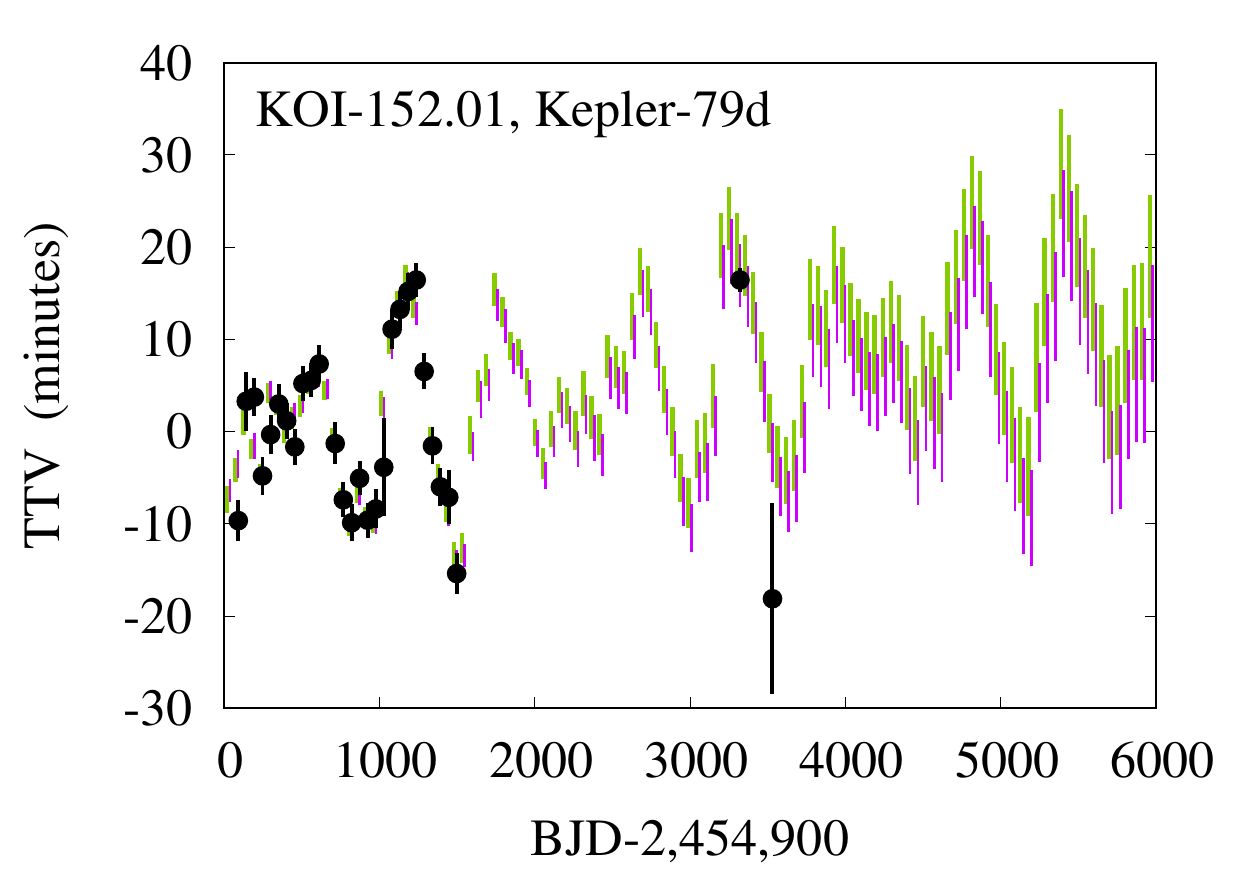}
	\includegraphics[width=0.46\linewidth]{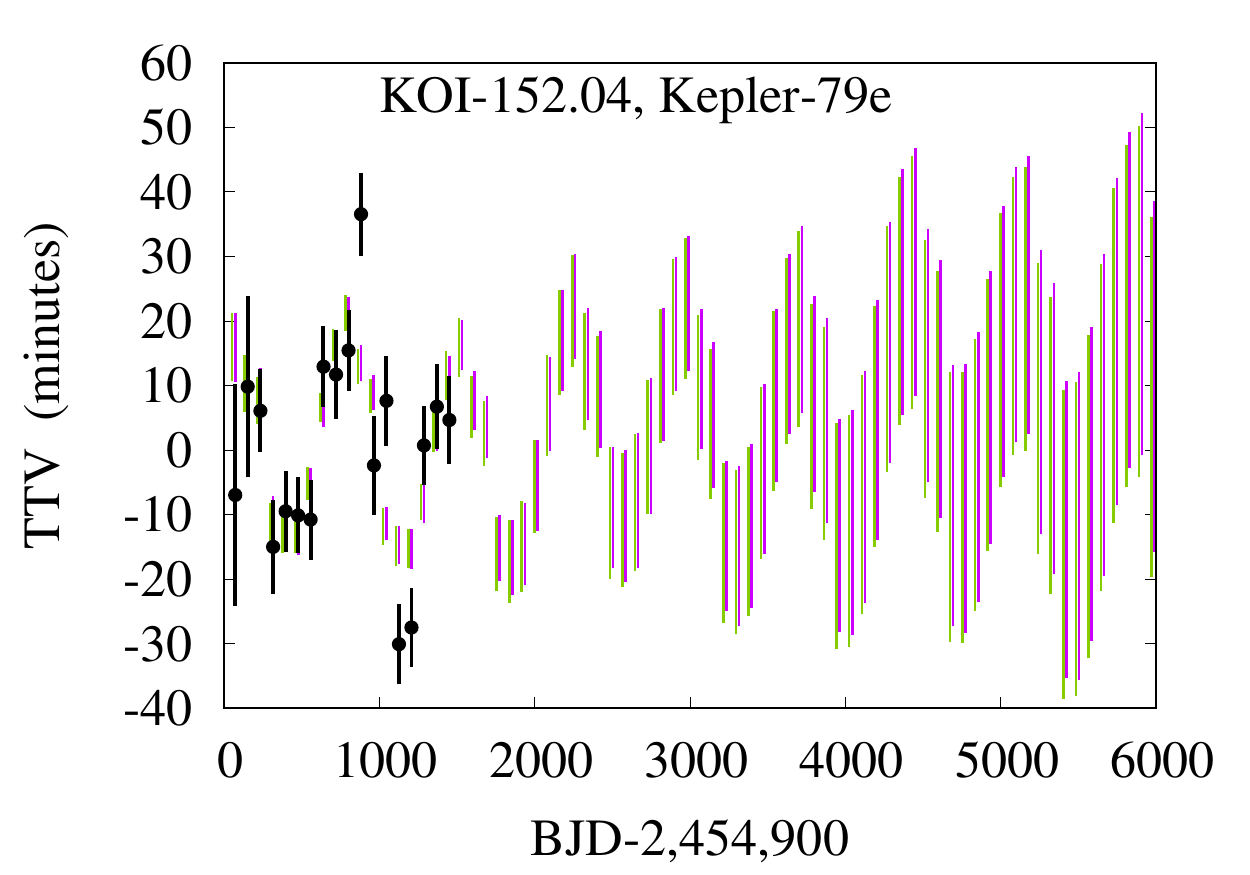}
	\caption{Observed and simulated deviations of transit times from a linear fit to the observed \emph{Kepler} data. The colored bands mark the standard deviation of 1000 simulated transit times from the posterior sampling models, with green marking the solutions following \emph{Kepler} only, and magenta marking the dataset including the \emph{HST} times.}
	\label{fig:ttv_models}
\end{figure*}

\subsection{Transit Timing Variation Fits} 
\label{sec:ttv}

Our transit timing dataset includes the \emph{Kepler} data analyzed by \cite{Jontof-Hutter2014}, who performed dynamical fits to the first 16 quarters of \emph{Kepler} data using all short cadence data available. To this dataset, we add the few Q 17 transits catalogued from long cadence data by \cite{Rowe2015}, and the two transit times measured from WFC3. The timing uncertainty on the first \emph{HST} visit was 1.31 minutes. During the second visit, data gaps during both ingress and egress reduced the precision on the transit timing, leaving an uncertainty of 10.4 minutes.

Our TTV models assumed coplanarity, and included five free parameters per planet: orbital period $P$, phase-at-epoch T$_0$, planet-star mass ratio $M_{\mathrm{p}} / M_*$, and the eccentricity vector components $e$ sin~$\omega$ and $e$ cos~$\omega$. Coplanarity is a reasonable assumption as mutual inclinations have little effect on transit times unless the mutual inclinations reach relatively large angles \citep{Nesvorny2014}, which is very unlikely for a multi-planet system where all planets transit the host star \citep[\emph{Kepler} multi-planet systems are nearly coplanar;][]{Fabrycky2014}. For orbital period, phase-at-epoch, and the planet-star mass ratio, we adopted a uniform prior. The planet-star mass ratio was also assumed to be positive definite as some of the posterior sampling reaches zero mass. No other limits are placed on these three parameters although in practice the orbital period and transit epoch are known so precisely that the samples never deviate from the best-fit value by more than a few minutes. For the eccentricity vector components $e$ sin~$\omega$ and $e$ cos~$\omega$, we assumed a Gaussian prior centered at zero with a standard deviation of 0.1. This Gaussian prior for eccentricity is motivated by the fact that high eccentricity solutions, which can often fit the data as well as the low eccentricity models, are unlikely or unstable for closely-packed multi-planet systems \citep[e.g.][]{Jontof-Hutter2015, Jontof-Hutter2016, Jontof-Hutter2019}. To sample the posteriors of these parameters, we used a Differential Evolution Markov Chain Monte Carlo algorithm \citep{Braak2006, Jontof-Hutter2015, Jontof-Hutter2016}, beginning the chains close to the best-fit model found by \cite{Jontof-Hutter2014}.

We perform a TTV model fit to just the \emph{Kepler} data as well as the full set of \emph{Kepler} + \emph{HST} data and compare the effect of adding the \emph{HST} data. Figure~\ref{fig:ttv_models} shows the observed and simulated TTVs for these data. The mid-transit time of the first visit with WFC is very close to the value predicted following the \emph{Kepler} data, while the mid-transit time of the second visit with the larger uncertainty is earlier than expected at the $\sim 1.5 \sigma$ level. The agreement between the predicted and observed transit time of the first \emph{HST} visit bolsters the mass measurements of the \emph{Kepler} only dataset, and confirms the low density of Kepler-79 d. The earlier time for the second visit, causes the TTV model to favor a slightly shorter orbital period for Kepler-79 d, although well within the uncertainty following the \emph{Kepler} dataset alone. The effect is a systematic revision of predicted transit times to occur several minutes earlier for the next few years. For the other planets, the \emph{HST} data has a smaller effect on the predicted transit times.

Table~\ref{table:ttv_fits} shows the parameter estimates from these fits. Although our mass estimates from a fit to the \emph{Kepler} data alone agree with the masses from \cite{Jontof-Hutter2014} to better than $ 1 \sigma$, they are consistently lower and possess smaller uncertainties than the previously published values. This is because the previous study utilized Levenberg-Marquardt $\chi^2$ minimization and estimated parameter values and uncertainties from a union of solutions found by this technique. This led to different median values for the planet masses and overestimated uncertainties. The results presented in this work are more appropriately derived from MCMC sampling of the posterior. Since the updated TTV values for Kepler-79d's period, eccentricity, and argument of periastron values are slightly different from those assumed in our light curve fitting (we used values from \citealp{Jontof-Hutter2014}, see \S~\ref{sec:lc_fits}), we refit our combined \emph{Kepler} and \emph{HST} light curves by placing TTV-derived Gaussian priors and accounting for covariances for these parameters. Results from this fit are in good agreement with the reported parameter estimates in \S~\ref{sec:wlc} (differences $\lesssim 1 \sigma$). We also use this fit's results to estimate the stellar mass \citep[e.g.][]{Winn2010} using the \emph{Gaia} stellar radius value and find that it is consistent with stellar mass value reported in \cite{Fulton2018}.

The addition of the new \emph{HST} data has a relatively minor effect on the best-fit TTV masses and corresponding uncertainties. This is because the \emph{Kepler} dataset already sampled many frequencies (near resonances as well as synodic chopping) in the TTVs with a high signal-to-noise ratio (SNR). This breaks the mass-eccentricity and eccentricity-eccentricity degeneracies for the Kepler-79 system \citep{Jontof-Hutter2019}, reducing the value of additional transit timing measurements at later epochs. Nonetheless, we find that the revision of the planetary masses to lower values, in particular for Kepler-79d, encourages continued interest in this planetary system and further cements Kepler-79d's status as a super-puff.

\subsection{Transmission Spectrum and Atmospheric Modeling}   \label{sec:transmission_spectrum}

\subsubsection{HST WFC3 Spectrum}
\begin{figure}
	\centering
	\includegraphics[width=\linewidth]{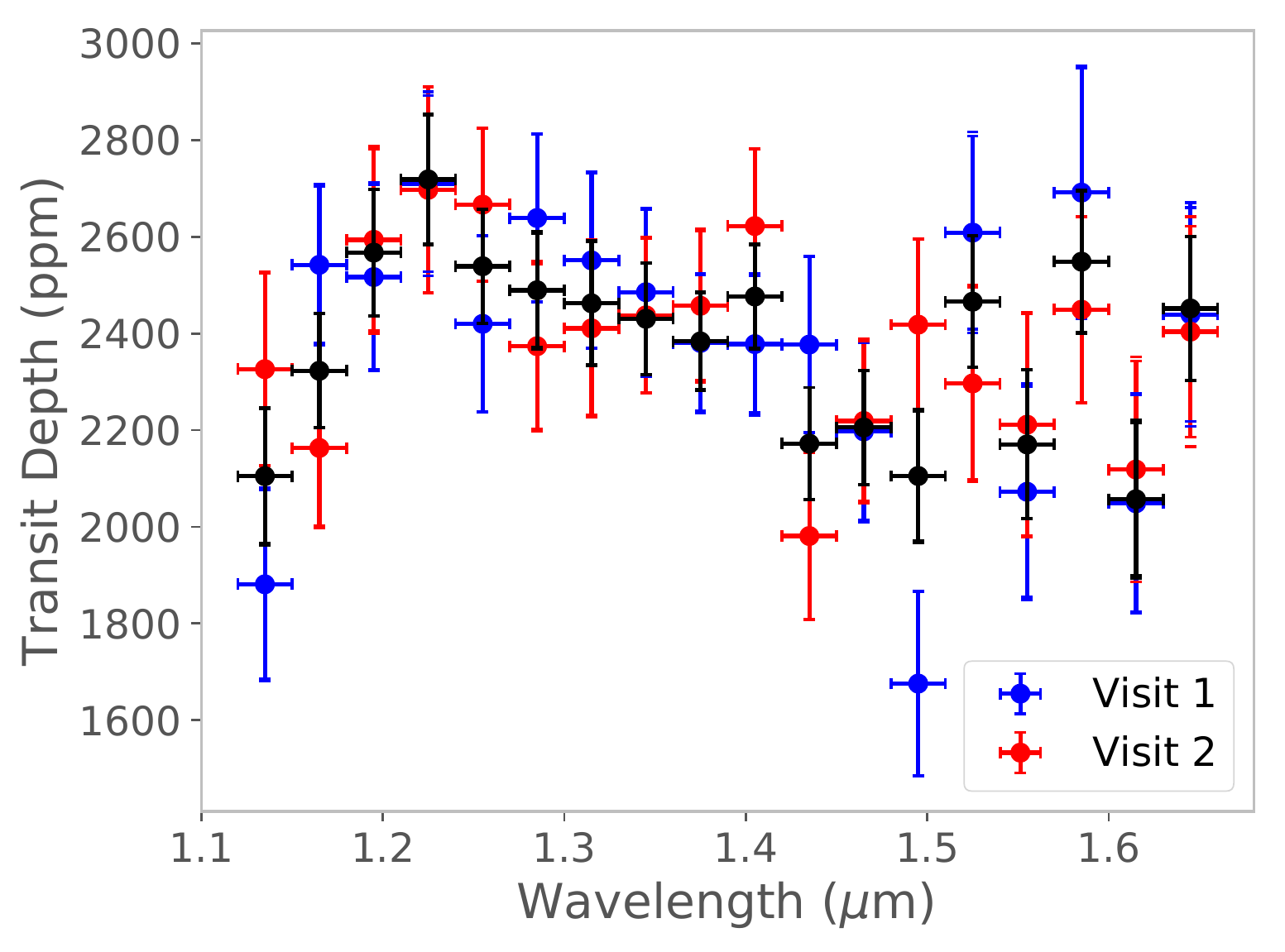}
	\caption{The transmission spectrum of Kepler-79d measured during 2 visits with the \emph{HST} WFC3 instrument. Black points show the spectrum obtained from a joint fit of the two visits. There is good agreement between the two visits except in two bandpasses centered at $1.135$ and $1.495 \mu$m, which show a larger scatter in the measured transit depths.}
	\label{fig:hst_wfc3_spectra}
\end{figure}

\begin{figure}
    \centering
    \includegraphics[width=\linewidth]{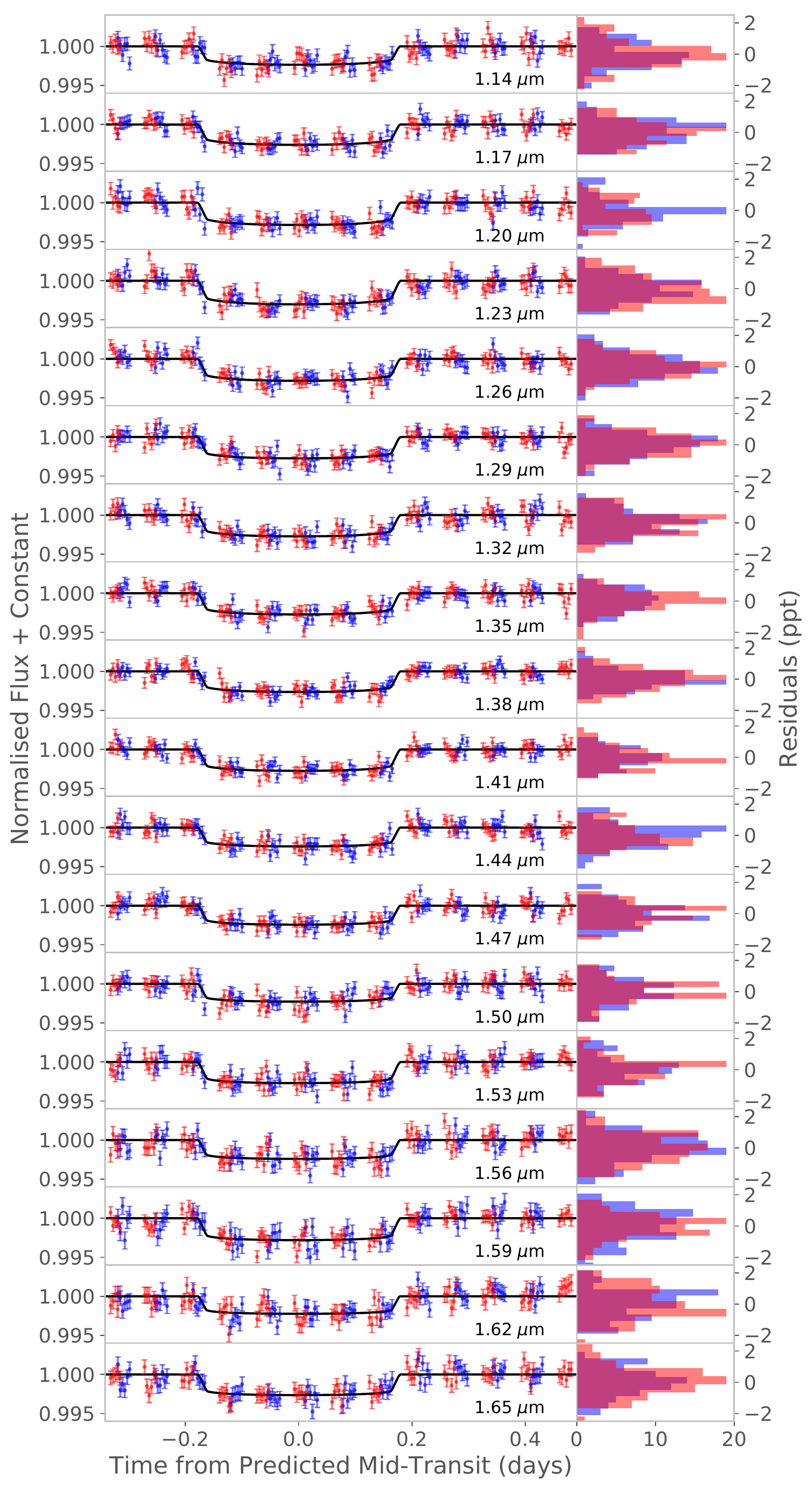}
    \caption{Spectroscopic light curves as well as the best-fit transit models for the 2 \emph{HST} visits (visit 1 in blue and visit 2 in red) and a histogram of the residuals (in parts per thousand).}
    \label{fig:spec_lc}
\end{figure}

The transmission spectrum for Kepler-79d in the \emph{HST} WFC3 bandpass is shown in Figure~\ref{fig:hst_wfc3_spectra}, with the corresponding $R_p / R_*$ values tabulated in Table~\ref{table:spectroscopic_fits}. There is good agreement between the spectra obtained from the two visits in all but two bandpasses. Transit depths measurements from the two visits in the $1.12 - 1.15$ $\mu$m and $1.48 - 1.51$ $\mu$m wavelength bands show a larger scatter. Spectroscopic light curves for both visits are shown for comparison in Figure~\ref{fig:spec_lc}. The transmission spectrum from the second visit has a smaller scatter in the spectroscopic transit depths and is more commensurate with a flat line than the spectrum from the first visit.

\begin{table}
    \centering
    \caption{Spectroscopic Light Curve Fit Results}
    \begin{tabular*}{\linewidth}{@{\extracolsep{\fill}} lcccc}
    \hline \hline
    Wavelength & $R_p / R_*$ & $\pm 1 \sigma$ & Transit Depth & $\pm 1 \sigma$ \\ 
    ($\mu$m) &  &  & (ppm) & (ppm) \\ \hline 
    1.120-1.150 & 0.04588 & 0.00154 & 2105 & 141 \\
    1.150-1.180 & 0.04820 & 0.00122 & 2323 & 118 \\
    1.180-1.210 & 0.05067 & 0.00129 & 2567 & 131 \\
    1.210-1.240 & 0.05214 & 0.00129 & 2719 & 135 \\
    1.240-1.270 & 0.05039 & 0.00117 & 2539 & 118 \\
    1.270-1.300 & 0.04989 & 0.00120 & 2489 & 120 \\
    1.300-1.330 & 0.04962 & 0.00129 & 2462 & 128 \\
    1.330-1.360 & 0.04930 & 0.00117 & 2430 & 115 \\
    1.360-1.390 & 0.04882 & 0.00103 & 2383 & 101 \\
    1.390-1.420 & 0.04976 & 0.00108 & 2476 & 108 \\
    1.420-1.450 & 0.04661 & 0.00124 & 2172 & 116 \\
    1.450-1.480 & 0.04696 & 0.00126 & 2206 & 118 \\
    1.480-1.510 & 0.04588 & 0.00148 & 2105 & 136 \\
    1.510-1.540 & 0.04966 & 0.00137 & 2466 & 136 \\
    1.540-1.570 & 0.04659 & 0.00165 & 2170 & 154 \\
    1.570-1.600 & 0.05048 & 0.00146 & 2548 & 147 \\
    1.600-1.630 & 0.04535 & 0.00177 & 2057 & 161 \\
    1.630-1.660 & 0.04951 & 0.00150 & 2452 & 149 \\
    \hline
	\end{tabular*}
	\label{table:spectroscopic_fits}
\end{table}

Table~\ref{tab:hst_evidence} lists the values of Bayesian evidence for the fiducial constant and linear models as well as some physically plausible models that represent limiting cases for Kepler-79d's atmosphere. We calculate the Bayes factor $\mathcal{B}$ (ratio of evidence) for a particular model by comparing its evidence with that of the constant transit depth model. We find that the HST data provide moderate evidence ($\mathcal{B}$ of 16.9) in favor of the constant model relative to the model with a linear trend in the WFC3 transmission spectrum. We quantify the statistical significance of the rise in transit depth between $1.1 \mu$m and $1.2 \mu$m as well as the dip around $1.45 \mu$m by comparing the evidence for the constant transit depth model with a squared exponential GP model. The GP model provides us with a non-parametric way that is independent of any forward model for fitting the shape of the transmission spectrum. With a $\mathcal{B}$ of 3.1 in favor of the GP model, the structure in the spectrum is only marginally significant.

We next investigate a range of physically motivated models in order to determine which of these models are ruled out by our data.  Given Kepler-79b's low density, it is reasonable to consider very low metallicity atmospheric compositions. We therefore compare the data with forward models that only includes collision induced absorption (CIA) and Rayleigh scattering contributions from hydrogen and helium \citep[see also][]{Libby-Roberts2020}. We use the Planetary Atmospheric Transmission for Observer Noobs (\texttt{PLATON}) atmospheric modeling and retrieval suite \citep{Zhang2019, Zhang2020} to compare the data and models in a retrieval framework. For all of our retrievals with \texttt{PLATON}, we place Gaussian priors on planet mass and stellar radius. For the metal-free atmosphere scenario, we initially allow both the planet's radius (at a pressure of 1 bar) and the atmosphere's isothermal temperature to vary. We find that the temperature is poorly constrained and its posterior covers the entire prior range ($200 - 700$ K), with a marginal preference for lower temperatures. Fixing the temperature to $630$ K and fitting only for the planet radius only results in a slight decrease in the model evidence, as expected from the weak constraints on temperature in the previous fit. The constant transit depth model is moderately favored ($\mathcal{B}$ of $2.9-8$) over either of these models.

\begin{figure*}
	\centering
	\includegraphics[width=\linewidth]{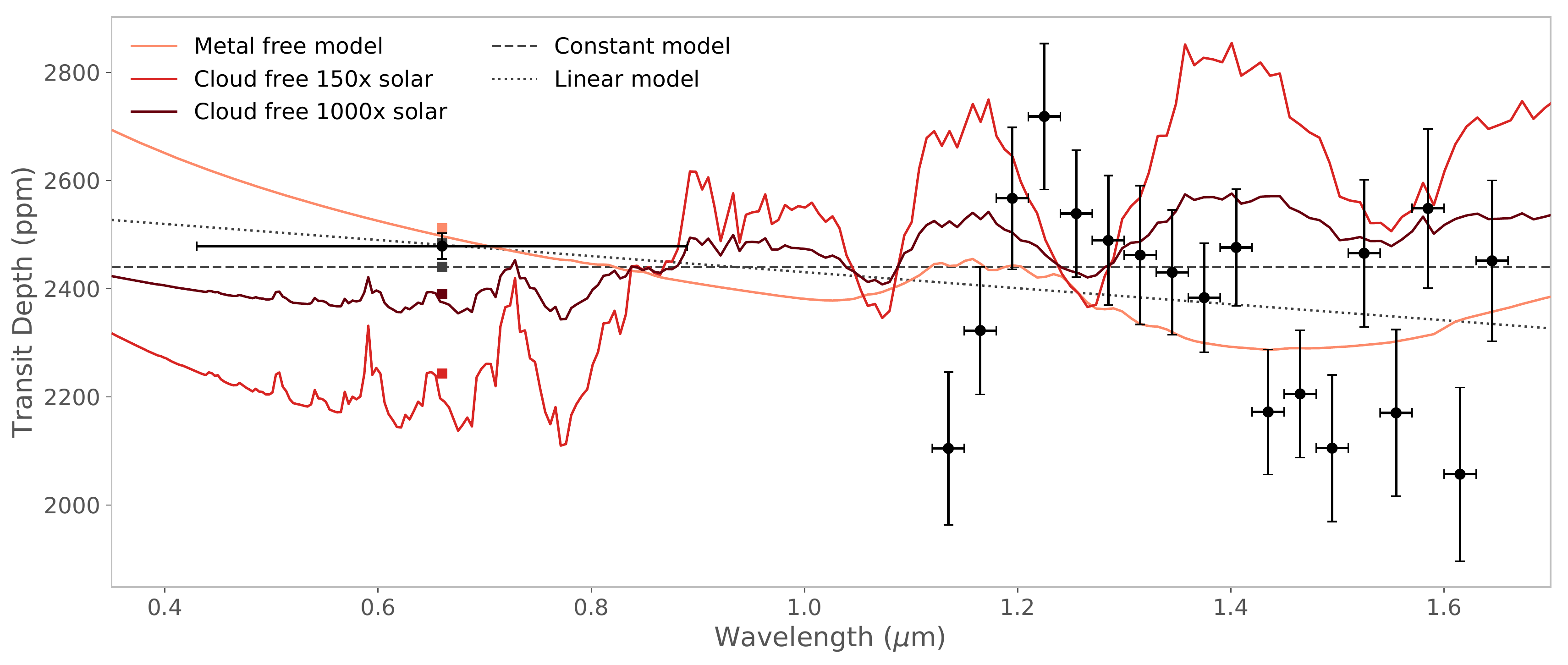}
	\caption{The transmission spectrum of Kepler-79d measured with \emph{Kepler} and \emph{HST} WFC3. The best-fit cloud free models with metal poor composition, $150 \times$ solar metallicity, and $1000 \times$ solar metallicity are plotted along with the best-fit constant and linear models. The data are consistent with a constant transit depth model. The Bayesian evidence and the Bayes factor for these models are reported in Table~\ref{tab:kepler_hst_evidence}.}
	\label{fig:trans_spec}
\end{figure*}

\begin{table}
    \centering
    \caption{\emph{HST} Model Evidence}
    \begin{tabular*}{\linewidth}{@{\extracolsep{\fill}} lcc}
    \hline \hline
    Model & log$_e$ (Evidence) & Bayes factor ($\mathcal{B}$) \\ \hline 
    Constant & 124.6 & -- \\
    Linear & 121.8 & 1:16.9  \\
    $150 \times$ solar (fixed $T$) & 116.8 & 1:2578  \\
    $1000 \times$ solar (fixed $T$) & 121.6 & 1:19.9 \\
    Metal free (fixed $T$) & 122.5 & 1:8 \\
    Metal free & 123.6 & 1:2.9 \\
    \hline
	\end{tabular*}
	\label{tab:hst_evidence}
\end{table}

In principle, this planet's low measured density implies a strict upper limit of $150 \times$ solar on its bulk metallicity and consequently its atmospheric metallicity \citep[assuming the metals and envelope are homogeneously mixed throughout;][]{Lopez2014, Thorngren2019}\footnote{Although we provide updated mass and radius measurements in this work, we expect that these updates will not significantly affect the bulk metallicity estimate reported in in \cite{Lopez2014}.}. However, this upper limit can be relaxed to values as high as $350 \times$ solar if high-altitude hazes shift the photospheric pressure to $\sim 10 \mu$m bar or $1000 \times$ solar if tidal heating \citep[e.g. from obliquity tides,][]{Millholland2019} augments the internal heat flux of the planet. Both the presence of hazes at low pressures and the increase in internal heat flux reduce the envelope to core mass ratio required to match the planet's mass and radius, thereby increasing the upper limit of the planet's bulk and atmospheric metallicity. We therefore also consider two higher metallicity atmosphere models, where we fix the metallicity to either $150 \times $ or $1000 \times$ solar metallicity.  As before, we fix the atmospheric temperature to 630 K.  We also assume a solar C/O ratio $= 0.53$, include Rayleigh scattering from gas, and exclude clouds and any other sources of scattering. The only quantity that we vary is then the planet's radius. The constant transit depth model is strongly favored over either the $150 \times $ and $1000 \times$ solar models with $\mathcal{B}$ of $2578$ and $19.9$ respectively.

\subsubsection{Fitting the WFC3 spectrum and Kepler WLC depth}

The relative values of the \emph{Kepler} white light curve depth and the WFC3 transmission spectrum give us important information about Kepler-79d's atmosphere, especially if scattering from aerosols dominates the absorption cross-section. Kepler-79d is the first super-puff for which we can make such a comparison, as the optical transit depths for the Kepler-51 planets are strongly biased by stellar activity \citep{Libby-Roberts2020}. Table~\ref{tab:kepler_hst_evidence} lists the Bayesian evidence and the Bayes factor $\mathcal{B}$ relative to the constant transit depth model, where we have updated our fits to include both the \emph{Kepler} and \emph{HST} data. Figure~\ref{fig:trans_spec} shows Kepler-79d's transmission spectrum as well as the best-fit retrieved models for the different atmospheric scenarios listed in Table~\ref{tab:kepler_hst_evidence} and discussed below. We find that even with the addition of the \emph{Kepler} transit depth, the transmission spectrum prefers a constant transit depth model. Although the transit depths in the two bands differ by 100 ppm, this difference is only marginally significant ($2.2\sigma$) and therefore has a negligible influence on $\mathcal{B}$ for the linear model.

\begin{table}
    \centering
    \caption{\emph{Kepler} + \emph{HST} Model Evidence}
    \begin{tabular*}{\linewidth}{@{\extracolsep{\fill}} lcc}
    \hline \hline
    Model & log$_e$ (Evidence) & Bayes factor ($\mathcal{B}$) \\ \hline
    Constant & 131.8 & -- \\
    Linear & 128.9 & 1:16.7  \\
    $150 \times$ solar (fixed $T$) & 113.7 & $1:6.9 \times 10^7$  \\
    $1000 \times$ solar (fixed $T$) & 124.7 & 1:1157 \\
    Metal free (fixed $T$) & 124.4 & 1:1622 \\
    Metal free & 129.6 & 1:8.8 \\
    \hline
	\end{tabular*}
	\label{tab:kepler_hst_evidence}
\end{table}

The addition of the \emph{Kepler} depth has a larger influence on $\mathcal{B}$ for the metal poor model. The \emph{Kepler} depth provides a much stronger constraint on the Rayleigh scattering contribution of the hydrogen-helium atmosphere, which depends directly on the planet's scale height ($\mathrm{d} \; R_{\mathrm{p}} / \mathrm{d \; ln } \lambda = - 4 H$ in the Rayleigh regime where $H$ is the scale height). This leads to a strong preference for models with low atmospheric temperatures ($< 364$ K at $2 \sigma$ confidence) as the scale height for the zero-albedo full heat redistribution equilibrium temperature of $630$ K is much too large to fit the relative difference between the \emph{Kepler} and \emph{HST} transit depths. Fixing the temperature to $630$ K leads to a significant increase in the preferred planet mass ($7.1 \pm 0.7 \; \mathrm{M}_{\oplus}$), an increase that is driven by the need for a smaller scale height to match the transmission spectrum. The Gaussian prior we placed on the planet mass penalizes the model evidence for this increase and leads to a significant increase in $\mathcal{B}$. A metal free atmosphere with a plausible temperature of $\sim 600$ K at a slant optical depth of unity (at $\sim 0.1$ bar) is therefore ruled out by the data.

Although significantly cooler metal poor models provide an improved match to the data, a reduction of the temperature at this pressure by a factor of two is very unlikely: it would require highly inefficient heat redistribution and/or a very high albedo (comparable to that of the icy moons). Both theoretical and observation constraints favor efficient circulation and small day-night temperature gradients for planets cooler than $1000$ K \citep{Komacek2016, Cowan2011, Perez-Becker2013a, Garhart2020}. For planets with volatile-rich envelopes, the most plausible way to increase the albedo is to introduce an optically thick reflective cloud layer at low pressures. Not only are most clouds not expected to have such a high albedo, the presence of such a cloud layer would be inconsistent with the assumption of a very low atmospheric metallicity itself as metals are needed to form clouds. We therefore regard the metal poor model as a highly improbable explanation for the measured transmission spectrum.

The constant model is strongly favored over either of the cloud-free high atmospheric metallicity ($150 \times$ and $1000 \times$ solar) models when we include the Kepler depth. We conclude that there is no evidence for any of the expected absorption or scattering features in the combined \emph{Kepler} and \emph{HST} data.  This suggests that Kepler-79d must host a high-altitude cloud or haze layer that effectively mutes the signature of atmospheric absorption. The lack of a detectable scattering slope in the \emph{Kepler} and WFC3 bandpasses can be used to place a lower limit on the particle size distribution for this scattering haze. Rather than fitting a parameterized cloud model, we instead investigate whether microphysical models of photochemical hazes can match our observations. Although these same microphysical models can also be used to study condensate cloud formation, \cite{Gao2020} found that photochemical hazes dominate the scattering opacity for planets with temperatures similar to that of Kepler-79d.

\begin{figure}
	\centering
	\includegraphics[width=\linewidth]{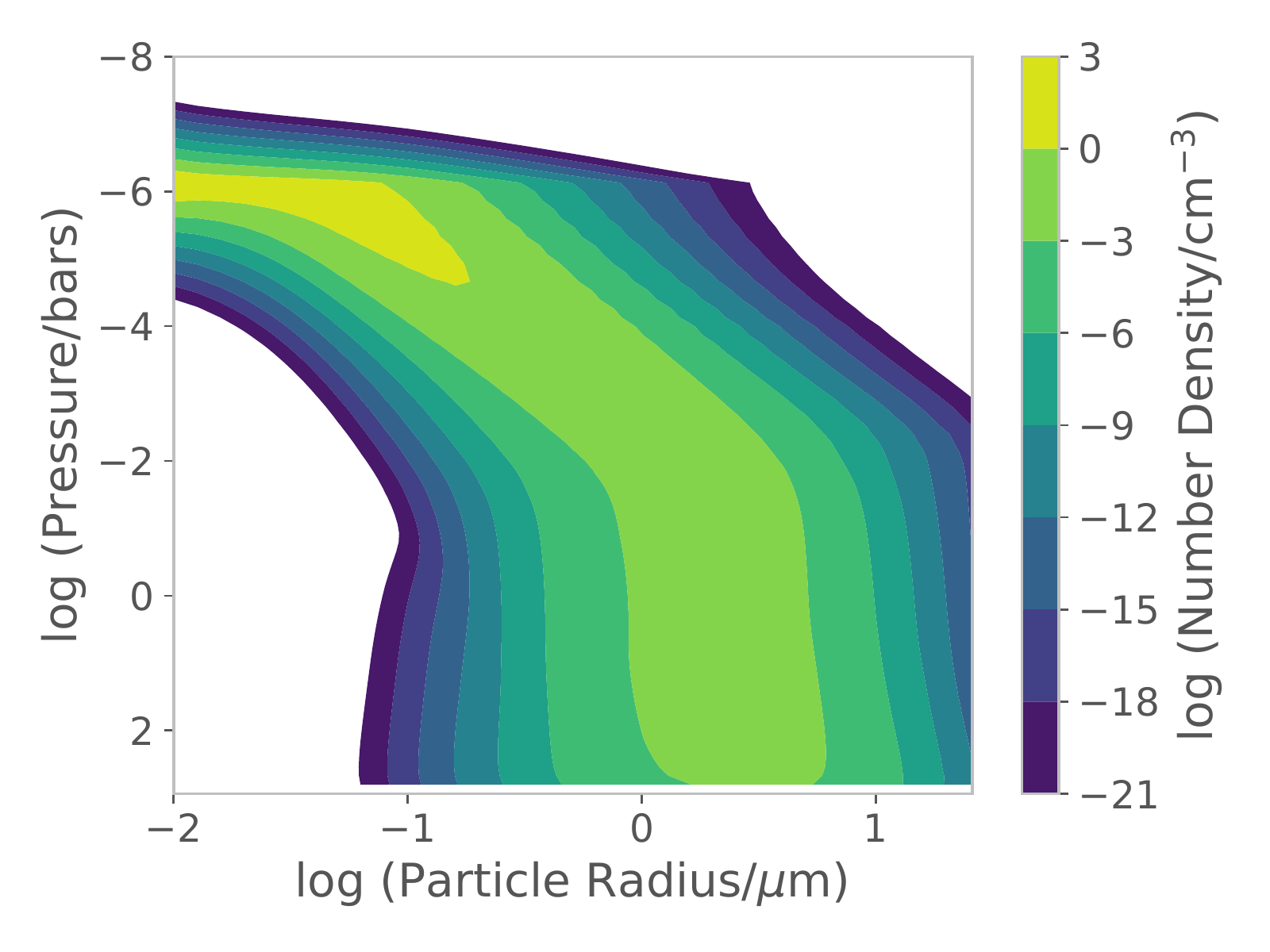}
	\caption{Number density of haze particles of different radii at different pressures levels in the atmosphere for $M_{\mathrm{core}} = 5 M_{\oplus}$, $T_{\mathrm{int}} = 75$ K, $K_{zz} = 10^{7}$ cm$^2$ s$^{-1}$, and atmospheric mass fraction of 18\%. Haze formation at low pressures and transport due to outflowing wind, vertical mixing, and sedimentation leads to an abundance of sub-micron sized particles at low pressures ($\sim 1 - 10 \mu$bar).}
	\label{fig:num_density_dist}
\end{figure}

\subsubsection{CARMA Photochemical Haze Models}
\label{sec:carma}

\begin{figure*}
	\centering
	\includegraphics[width=0.48\linewidth]{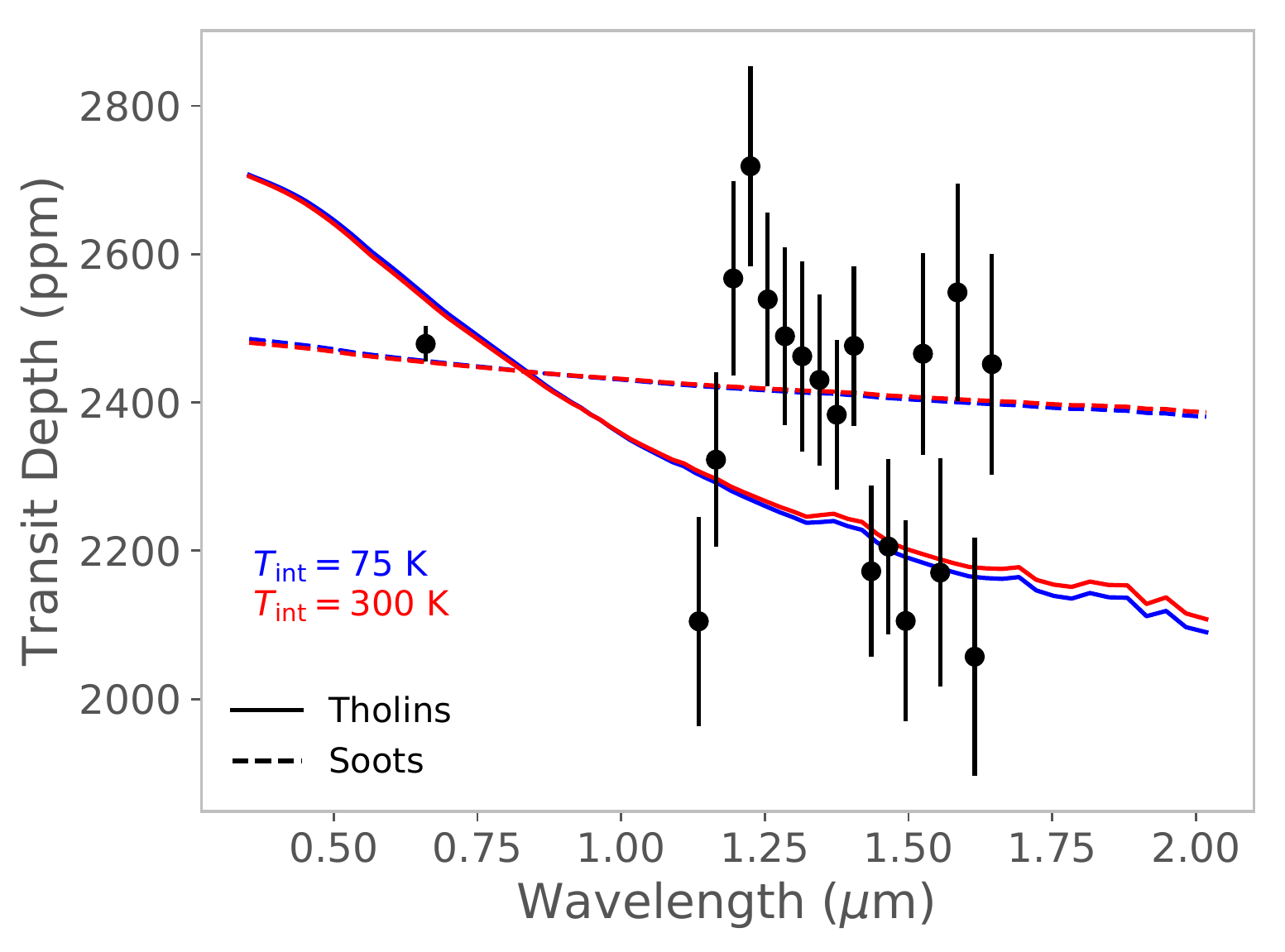}
	\includegraphics[width=0.48\linewidth]{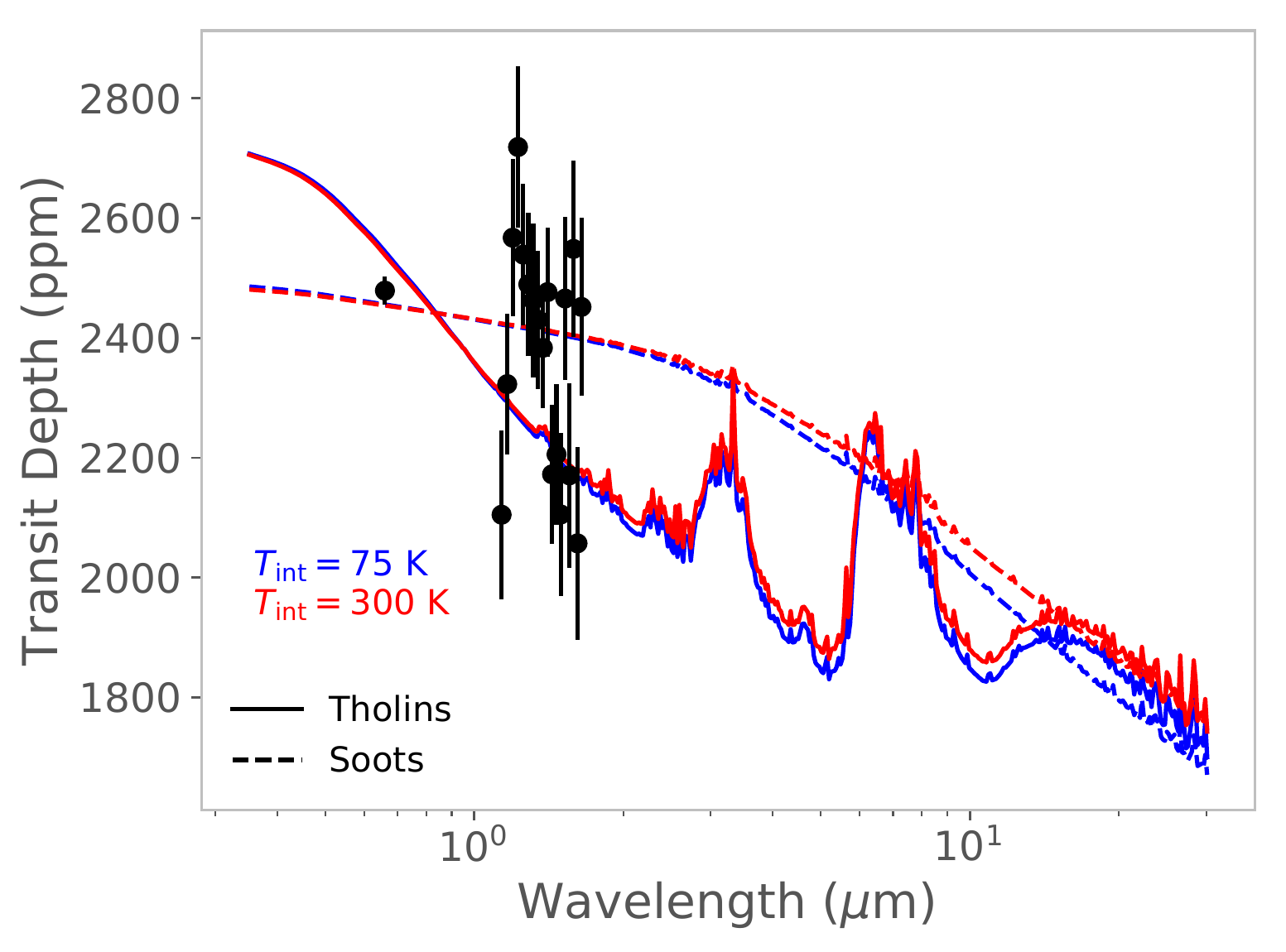}
	\caption{Forward CARMA models of the transmission spectrum of Kepler-79d for different optical properties of the aerosols (soots and tholins) and $T_{\mathrm{int}}$ values. In the left panel, we compare the models with the transmission spectra. The right panel shows the predicted transmission spectrum out to 30 $\mu$m.}
	\label{fig:carma_model_data}
\end{figure*}

In this section we investigate whether a physically motivated photochemical haze model can match the observed transmission spectrum. We use a modified and simplified version of the Community Aerosol and Radiation Model for Atmospheres (CARMA) to study the formation and distribution of hazes in Kepler-79d's atmosphere. CARMA is a 1D bin-scheme aerosol model that can account for microphysical processes such as nucleation, growth by condensation and coagulation, evaporation, and transport. The aerosol continuity equation is discretized over particle radius bins and mass exchange due to microphysical processes is allowed between these bins. For a more detailed description of CARMA, we direct the reader to \cite{Gao2018} and \cite{Adams2019}. Our modified and simplified CARMA model is fully described in \cite{Gao2020} and we briefly mention key features here for completeness.

Haze `seed' particles are generated at the pressure level where methane is photolyzed (typically centered around $\sim \mu$bar) with a production rate equal to the methane photolysis rate multiplied by an efficiency factor of 0.1. These seed particles are transported through the atmosphere and allowed to grow by coagulation. Particle transport usually includes sedimentation under the effect of gravity and turbulent vertical mixing. One additional important transport mechanism that is relevant to super-puffs is entrainment by the outflowing hydrodynamic wind that is responsible for mass loss. We take the outward flux due to this wind into account in our model and simplify other aspects of the model in order to keep it computationally tractable. The eddy diffusion coefficient ($K_{zz}$), which parameterizes vertical mixing, is assumed to be constant ($= 10^7$ cm$^2$ s$^{-1}$) throughout the atmosphere. The value of $K_{zz}$ is not well constrained by current observational data sets for transiting exoplanets, and a plausible range of values inferred from general circulation models can span many orders of magnitude \citep[e.g.][]{Moses2011, Parmentier2013, Charnay2015, Zhang2018b, Zhang2018c}. Following these works, we varied the value of this parameter between $10^6$ and $10^8$ cm$^2$ s$^{-1}$ and found that it had a negligible impact on our results, in agreement with \cite{Gao2020}. We additionally adopt a simple but adequate atmosphere model that incorporates both a convective and a radiative layer and uses the ideal gas law and hydrostatic equilibrium to obtain the atmospheric temperature-pressure (TP) profile (\citealp{Gao2020}, similar to previous models, e.g. \citealp{Owen2017}). 

The equilibrium temperature ($T_{\mathrm{eq}}$) of the planet is set to 630 K and we assume a core mass of 5 M$_{\oplus}$. For the internal heat flux of the planet, we choose values of $T_{\mathrm{int}}$ = 75 K and 300 K (where $F_{\mathrm{int}} = \sigma_{\mathrm{SB}} T_{\mathrm{int}}^4$). The former value reflects the expected residual heat of formation for a $\sim 1$ Gyr old planet with a mass equal to that of Kepler-79 d \citep{Lopez2014}. The latter value corresponds to the expected internal heat flux from dissipation due to obliquity tides \citep[][optimistic case, calculated using their Equation 11]{Millholland2019}. We only include hydrogen, helium, water, and methane (the primary constituents of a solar composition atmosphere at $\sim$600 K) in our models. All other species are not expected to contribute significantly either to the number density or the opacity in the optical and near infrared \citep{Lodders2002, Burrows2014}. Since the optical properties of the hazes are unknown, we consider two distinct end cases: scattering hazes (`tholins', refractive index taken from \citealp{Khare1984}) and absorbing hazes (`soots', \citealp{Morley2015, Lavvas2017}).

We find that the atmospheric mass fraction needed to fit the observed planet radius decreases when a high-altitude haze is present. The best-fit atmospheric mass fractions for $T_{\mathrm{int}} = 75$ K for soots and tholins are 17\% and 18\% respectively. These values are roughly half of the previous estimate, which used 20 mbar for the photospheric pressure \citep{Lopez2014}. Models containing soots require a slightly lower atmospheric mass fraction, as soot particles absorb more strongly than tholins. For $T_{\mathrm{int}} = 300$ K, \cite{Millholland2019} found that an atmospheric mass fraction of $\sim 10$\% could match the planetary radius. However, with haze formation, this value falls precipitously to $\sim 0.6$\%, which is akin to the typical atmospheric mass fraction for super-Earths in the \emph{Kepler} sample \citep[e.g.][]{Chiang2013, Lopez2014, Wolfgang2015, Owen2017}. This raises the intriguing possibility that some super-puffs might simply be super-Earths with unusually high internal heat fluxes and a high-altitude haze. However, whether such planets can manage to retain their atmospheres remains to be seen; we discuss this topic in more detail in \S~\ref{sec:puffy_mass_loss}.

Figure~\ref{fig:num_density_dist} shows the number density of haze particles for a representative CARMA model and Figure~\ref{fig:carma_model_data} shows the resulting transmission spectra for Kepler-79d. Figure~\ref{fig:num_density_dist} indicates that a large number of sub-micron particles are present at pressures of $1-10$ $\mu$bar, resulting in a transmission spectrum that is dominated by haze opacity. However, our predictions for the corresponding shape of the transmission spectrum can vary significantly depending on what we assume for the optical properties of the haze. We find that the soot models fit the data as well as our fiducial constant model ($\Delta$ BIC $\sim 1$ in favor of the soot models); our results are comparable in this case regardless of the value we assume for $T_{\mathrm{int}}$  (Figure~\ref{fig:carma_model_data}). Tholins provide a slightly worse fit ($\Delta$ BIC = 28 and 32 in favor of the constant model for $T_{\mathrm{int}}$ = 300 K and 75 K respectively). This suggests that that the prospective hazes in Kepler-79d's atmosphere are more absorbing at these wavelengths than their Titan counterparts. For Kepler-79d, CARMA provides a physically motivated forward model that is able to match the planet's transmission spectrum. It also offers predictions for the transmission spectrum at longer wavelengths that can be tested by facilities such as the \emph{James Webb Space Telescope} (\emph{JWST}) (Figure~\ref{fig:carma_model_data}, right panel) in the future.

\subsubsection{Stellar Activity}
\label{sec:stellar_activity}

In this section we investigate whether or not stellar activity might alter the shape of Kepler-79d's observed transmission spectrum at a level that would affect our interpretation of these data.  Kepler 79 is a $1.3^{+1.0}_{-0.4}$ Gyrs old quiescent late-F type star with a line-of-sight rotational velocity ($v$ sin~$i$) of $14 \pm 1$ km s$^{-1}$ \citep{Petigura2017a, Fulton2018}.  This corresponds to a rotation period $P \leq 4.8$ days, which is typical for main sequence \emph{Kepler} stars with effective temperatures within 2$\sigma$ of Kepler-79 (i.e. stars with $T_{\mathrm{eff}}$ in the range $ 6389 \pm 120$ K; \citealp{McQuillan2014}). We examined median normalized \emph{Kepler} light curves and found that this star's peak-to-peak variability is $< 0.2$\%, which is typical for F stars in the \emph{Kepler} sample ($0.13^{+0.23}_{-0.06}$\%; \citealp{McQuillan2014, Rackham2019}). Such a low variability amplitude would cause the measured transit depth to vary by less than 5 ppm from one epoch to the next in the \emph{Kepler} band; this is much smaller than the measurement errors for individual \emph{Kepler} transit depths. We see no evidence for any spot crossings in the \emph{Kepler} (and WFC3) transits of Kepler-79d, which has the deepest and best signal-to-noise ratio amongst the four planets. For the other three planets, which have significantly shallower transits, the signal-to-noise ratio for individual transits is too low to provide useful constraints on spot or faculae crossings. We find no evidence for epoch-to-epoch variability in the measured \emph{Kepler} transit depths for Kepler-79d (see \S~\ref{sec:wlc}), in good agreement with the lack of detectable photometric variability and the apparent absence of any spot or faculae occultations.

Although we can place a tight upper limit on Kepler-79's photometric variability, unocculted spots and faculae that are nearly uniformly distributed in longitude may still introduce a wavelength dependent bias in the transit depths while maintaining a near-constant stellar flux. Spots could remain unocculted if they occur at a specific range of latitudes (similar to the Sun) that the transiting planet does not traverse; this is plausible for the case where the planet's orbit is well-aligned with the star's spin axis, but unlikely for planets with nearly pole-on orbits. Although the 100 ppm offset between the measured \emph{Kepler} and \emph{HST} transit depths for Kepler-79d is only marginally significant ($2.2 \sigma$; see \S~\ref{sec:wlc} \& \ref{sec:transmission_spectrum}), we can nonetheless use it to place limits on the spot properties of Kepler-79. Regardless of whether or not this offset is produced by unocculted spots, it gives us a useful metric for what might be considered a significant effect of the star spots on the transmission spectrum of Kepler-79d.

We estimate the fractional area of the star that must be covered by spots (spot coverage fraction $\varepsilon$) for a range of spot temperatures in order to reproduce the $\sim$100 ppm offset between the \emph{Kepler} and \emph{HST} depths. To do this, we follow the procedure outlined in \cite{Evans2018} and fit our transmission spectrum assuming that the planet has the same underlying transit depth $D$ in this wavelength range ($0.4 - 1.7 \mu$m), which is then  altered by the spots in a wavelength dependent manner to produce the observed depth $D_{\mathrm{obs}, \lambda}$:
\begin{equation}
    D_{\mathrm{obs}, \lambda} = \frac{D}{1 - \varepsilon (1 - F_{\mathrm{spot}, \lambda} / F_{*, \lambda})} = \frac{D}{1 - \varepsilon \alpha}
\end{equation}
where $F_{\mathrm{spot}, \lambda}$ and $F_{*, \lambda}$ are the stellar intensity profiles (in this  case, BT-NextGen stellar models; \citealp{Allard2012}) for the temperatures corresponding to the spots and the homogeneous stellar surface respectively, and $\alpha$ is the resulting spot contrast. We fix the stellar surface's temperature to $6389$ K and consider spot temperatures that decrease in increments of 500 K up to a temperature difference of 2500 K (i.e. minimum spot temperature of 3889 K). As the spot temperature decreases, the best-fit value of $D$ decreases from $2267$ ppm to $2108$ ppm and the best-fit $\varepsilon$ decreases from 30\% (for $T_{\mathrm{spot}} = 5889$ K) to $\sim 15$\% (for $T_{\mathrm{spot}} \lesssim 4889$ K). 

We next ask whether such values for $\varepsilon$ are compatible with our upper limit on the observed photometric variability of Kepler-79 in the \emph{Kepler} bandpass. Unfortunately, the predicted photometric variability is degenerate with the assumed spot properties including temperature and size. Nonetheless, for random distribution of spots in longitude, we can expect the variability amplitude to scale as $\alpha \Omega \sqrt{n}$, where $\Omega$ is the solid angle of a spot and $n$ is the number of spots \citep{Rackham2019}. To pin this relationship to an absolute value of the variability amplitude, we use the \emph{Kepler} variability amplitude determined by \cite{Rackham2019} for a specific spot temperature and size. For a spot temperature of $\sim 4290$ K (based on scaling from spot temperatures for stars in the spectral range G1$-$M3), spot size of $2^{\circ}$ (covering $100$ ppm of the entire stellar surface; based on observations of large spot groups on the Sun), and $\varepsilon = 15\%$ (required by our fits to produce a $100$ ppm offset between the \emph{Kepler} and \emph{HST} bandpasses), Kepler-79 should display variability in the \emph{Kepler} bandpass with an amplitude of $1.8^{+0.8}_{-0.6} \%$. This value is an order of magnitude larger than the upper limit of $<0.2\%$ that is inferred from \emph{Kepler} observations of Kepler-79.

Keeping the spot size the same and varying only the spot temperature (and consequently $\alpha$), we find that an order of magnitude decrease in variability amplitude requires a spot temperature of $\sim 6239$ K (150 K cooler than the stellar photosphere) and an implausibly large $\varepsilon = 82\%$. This value of $\varepsilon$ is large enough that it would significantly affect the spectral characterization of Kepler-79. Conversely, for a fixed spot temperature of $\sim 4389$ K and fixed $\varepsilon = 15\%$, the spots' $\Omega$ would need to be roughly two orders of magnitude smaller (corresponding to $\sim 1$ ppm of the stellar surface) to be consistent with our upper limit on the observed variability amplitude. The reduction in $\Omega$ is roughly two orders of magnitude rather than one because reducing $\Omega$ while keeping $\varepsilon$ fixed also leads to an increase in the spot number $n$, which acts to offset the effect of reducing $\Omega$ on the variability amplitude. A spot covering $1$ ppm of the stellar surface would be comparable in size to a single granule on the surface of Kepler-79 (using scaling relationships from \citealp{Freytag1997, Trampedach2013, Tremblay2013}). We might expect that spots on Kepler-79 would be smaller than those on the Sun, as it rotates faster and has a thinner outer convective zone (spot area $ \propto \omega^{-1} \; (\rho_{\mathrm{b}} / \rho_\mathrm{t})$, where $\omega$ is the rotational frequency, and $\rho_{\mathrm{b}}$ and $\rho_\mathrm{t}$ are the densities at the base and the top of the outer convective zone; \citealp{Schmitt1983, Giampapa1984}) but spots this small are close to the physically plausible limit for Kepler-79, and therefore seem unlikely in practice. Even if they are present with a coverage fraction $\varepsilon$ of $15\%$, corresponding to the maximum coverage fraction consistent with the observed \emph{Kepler} variability amplitude, these spots would only produce a $100$ ppm (2.2 $\sigma$) offset between the measured \emph{Kepler} and \emph{HST} transit depths.

In the end, we find no evidence to suggest that stellar activity has appreciably altered the measured shape of Kepler-79b's transmission spectrum.  Furthermore, allowing for the potential presence of spots does not lead to a material change in our picture of Kepler-79d's atmosphere. If spots are present, they might marginally influence our inferences regarding the haze particle size distribution by changing the slope of the spectrum, but they cannot render the existence of haze particles unnecessary. These haze particles are required to explain the featureless WFC3 spectrum (\S~\ref{sec:transmission_spectrum}) and their presence at low pressures is also needed to reconcile Kepler-79b's predicted mass loss rate with its atmospheric lifetime (\S~\ref{sec:puffy_mass_loss}). Future observations at longer infrared wavelengths with \emph{JWST} will be even less sensitive to the potential presence of spots and will place much tighter constraints on the underlying particle size distribution in Kepler-79d's atmosphere.

\subsection{Implications of Mass Loss for Puffy Planets}    \label{sec:puffy_mass_loss}

The extended atmospheres and low surface gravities of super-puffs make them vulnerable to catastrophic mass loss. The question of how their inferred mass loss histories can be reconciled with their present-day ages is a matter of considerable debate. One promising idea put forward by \cite{Wang2019} and \citet{Gao2020} is that the measured radii of super-puffs are inflated by the presence of dust particles lofted by the atmospheric outflow ($\tau\sim$1 at tens of $\mu$bar, see Figure~\ref{fig:num_density_dist}).  If these planets are systematically smaller and denser than their observed radii would seem to suggest, it would significantly reduce the estimated mass loss rates. Here, we investigate whether or not this mechanism suffices to explain the observed properties of low-density planets from the \emph{Kepler} survey. We select a sample of systems from the TTV catalogue of \cite{Hadden2017} that have robust mass determinations (robust flag = 1) and updated stellar parameters from Gaia \citep{Fulton2018}. Since TTV fits constrain the planet to star mass ratios, we update the corresponding planet masses using the new stellar masses from Gaia\footnote{\cite{Fulton2018} note that the uncertainties on these stellar masses are likely to be underestimated as they are obtained by fitting isochrones. Nonetheless, the updated estimates are likely to be more accurate than the previously published values for these systems and we therefore adopt them in this study.}. We then select the subsample of planets with masses $< 100 \; \mathrm{M}_{\oplus}$ and bulk density $< 1$ g cm$^{-3}$. These limits are generous enough to ensure that we do not exclude any potential super-puffs from this sample. Table~\ref{table:planet_properties} shows some key properties for the planets in our sample.

\begin{table*}
	\begin{threeparttable}
    	\caption{Planet Properties for our `Puffy' Planets Sample} \label{table:planet_properties}
    	\begin{center}
            \begin{tabular*}{\linewidth}{@{\extracolsep{\fill}} lccccccc}
            \hline \hline
            Name & Mass\textsuperscript{a} & Radius\textsuperscript{b} & Density & Semimajor Axis\textsuperscript{b} & Incident Stellar & Age\textsuperscript{b} & log$_{10}$(Atmsopheric \\ 
             &  ($M_{\oplus}$) & ($R_{\oplus}$) & (g/cc) & (AU) &  Flux\textsuperscript{b} ($F_{\oplus}$) & (Gyrs) & Lifetime/Gyrs) \\ \hline 
            Kepler-9 b & $43.2^{+1.3}_{-1.2}$ & $8.1 \pm 0.2$ & $0.45^{+0.04}_{-0.03}$ & $0.1418 \pm 0.0012$ & $47.2 \pm 3.2$ & $1.8^{+1.5}_{-1.2}$ & $2.9^{+0.3}_{-0.6}$ \\
            Kepler-9 c & $29.7^{+0.9}_{-0.8}$ & $8.1 \pm 0.2$ & $0.31^{+0.03}_{-0.02}$ & $0.2266 \pm 0.0019$ & $18.5 \pm 1.2$ & $1.8^{+1.5}_{-1.2}$ & $2.9^{+0.3}_{-0.6}$ \\
            Kepler-11 e & $7.3^{+1.1}_{-1.1}$ & $4.0 \pm 0.1$ & $0.61^{+0.11}_{-0.10}$ & $0.1964 \pm 0.0022$ & $30.3 \pm 2.1$ & $6.5^{+1.9}_{-1.8}$ & $2.6^{+0.2}_{-0.2}$ \\
            Kepler-11 f & $1.9^{+0.5}_{-0.4}$ & $2.8 \pm 0.2$ & $0.46^{+0.17}_{-0.13}$ & $0.2526 \pm 0.0029$ & $18.3 \pm 1.3$ & $6.5^{+1.9}_{-1.8}$ & $-0.6^{+2.1}_{-1.8}$ \\
            Kepler-18 d & $14.8^{+2.7}_{-4.0}$ & $5.1 \pm 0.1$ & $0.60^{+0.13}_{-0.16}$ & $0.1177 \pm 0.0010$ & $46.5 \pm 3.2$ & $2.0^{+1.8}_{-1.7}$ & $2.1^{+0.4}_{-1.0}$ \\
            Kepler-33 d & $4.3^{+2.0}_{-2.0}$ & $4.5 \pm 0.1$ & $0.25^{+0.12}_{-0.12}$ & $0.1626 \pm 0.0022$ & $109.8 \pm 8.3$ & $4.8^{+1.4}_{-0.6}$ & $-0.5^{+3.3}_{-3.2}$ \\
            Kepler-33 e & $6.1^{+1.1}_{-1.0}$ & $3.5 \pm 0.1$ & $0.79^{+0.17}_{-0.15}$ & $0.2092 \pm 0.0028$ & $66.4 \pm 5.1$ & $4.8^{+1.4}_{-0.6}$ & $2.1^{+0.2}_{-0.2}$ \\
            Kepler-36 c & $7.7^{+0.3}_{-0.2}$ & $4.0 \pm 0.1$ & $0.68^{+0.08}_{-0.07}$ & $0.1269 \pm 0.0009$ & $190.3 \pm 12.7$ & $7.4^{+0.5}_{-0.5}$ & $2.0^{+0.1}_{-0.1}$ \\
            Kepler-33 f & $10.6^{+1.6}_{-1.5}$ & $3.9 \pm 0.1$ & $0.95^{+0.18}_{-0.16}$ & $0.2480 \pm 0.0033$ & $47.2 \pm 3.6$ & $4.8^{+1.4}_{-0.6}$ & $3.1^{+0.2}_{-0.2}$ \\
            Kepler-51 b\textsuperscript{c} & $3.5^{+1.8}_{-1.5}$ & $6.9 \pm 0.1$ & $0.06^{+0.03}_{-0.03}$ & $0.2423 \pm 0.0013$ & $10.7 \pm 0.8$ & $0.5^{+0.2}_{-0.2}$ & $-0.8^{+3.4}_{-2.9}$ \\
            Kepler-51 c\textsuperscript{c} & $4.2^{+0.5}_{-0.5}$ & $9.0 \pm 2.8$ & $0.03^{+0.06}_{-0.02}$ & $0.3702 \pm 0.0020$ & $4.6 \pm 0.3$ & $0.5^{+0.2}_{-0.2}$ & $0.8^{+0.7}_{-0.8}$ \\
            Kepler-51 d\textsuperscript{c} & $5.4^{+1.1}_{-1.0}$ & $9.5 \pm 0.2$ & $0.03^{+0.01}_{-0.01}$ & $0.4907 \pm 0.0026$ & $2.6 \pm 0.2$ & $0.5^{+0.2}_{-0.2}$ & $1.4^{+0.5}_{-0.7}$ \\
            Kepler-79 d\textsuperscript{d} & $5.3^{+0.9}_{-0.9}$ & $7.2 \pm 0.2$ & $0.08^{+0.02}_{-0.02}$ & $0.2937 \pm 0.0027$ & $30.0 \pm 2.1$ & $1.3^{+1.0}_{-0.4}$ & $0.4^{+1.5}_{-1.3}$ \\
            Kepler-79 e\textsuperscript{d} & $3.8^{+0.7}_{-0.6}$ & $3.5 \pm 0.2$ & $0.48^{+0.09}_{-0.10}$ & $0.3945 \pm 0.0037$ & $16.6 \pm 1.2$ & $1.3^{+1.0}_{-0.4}$ & $1.5^{+0.4}_{-0.4}$ \\
            Kepler-89 c & $8.8^{+3.0}_{-2.5}$ & $3.9 \pm 0.1$ & $0.84^{+0.30}_{-0.25}$ & $0.0986 \pm 0.0008$ & $252.2 \pm 15.9$ & $3.5^{+0.6}_{-0.6}$ & $1.5^{+0.3}_{-0.4}$ \\
            Kepler-89 d & $62.1^{+10.3}_{-10.8}$ & $10.1 \pm 1.6$ & $0.33^{+0.24}_{-0.13}$ & $0.1640 \pm 0.0013$ & $91.2 \pm 5.8$ & $3.5^{+0.6}_{-0.6}$ & $3.4^{+0.3}_{-0.3}$ \\
            Kepler-177 b & $5.0^{+0.8}_{-0.8}$ & $4.5 \pm 0.4$ & $0.30^{+0.11}_{-0.08}$ & $0.2110 \pm 0.0018$ & $38.1 \pm 3.5$ & $11.7^{+1.1}_{-1.0}$ & $2.4^{+0.2}_{-0.2}$ \\
            Kepler-177 c & $12.2^{+2.4}_{-2.3}$ & $9.3 \pm 0.4$ & $0.08^{+0.02}_{-0.02}$ & $0.2566 \pm 0.0022$ & $25.8 \pm 2.3$ & $11.7^{+1.1}_{-1.0}$ & $3.0^{+0.2}_{-0.2}$ \\
            Kepler-223 b & $3.7^{+1.8}_{-2.0}$ & $2.8 \pm 0.2$ & $0.88^{+0.54}_{-0.48}$ & $0.0752 \pm 0.0008$ & $414.4 \pm 57.1$ & $8.9^{+1.1}_{-1.0}$ & $-2.6^{+3.5}_{-3.6}$ \\
            Kepler-223 c & $12.1^{+2.6}_{-2.7}$ & $4.3 \pm 0.8$ & $0.79^{+0.69}_{-0.34}$ & $0.0912 \pm 0.0009$ & $281.3 \pm 38.2$ & $8.9^{+1.1}_{-1.0}$ & $2.4^{+0.4}_{-0.4}$ \\
            Kepler-223 d & $5.9^{+1.9}_{-1.8}$ & $5.9 \pm 0.8$ & $0.15^{+0.10}_{-0.06}$ & $0.1196 \pm 0.0012$ & $163.8 \pm 22.4$ & $8.9^{+1.1}_{-1.0}$ & $-0.6^{+2.5}_{-2.1}$ \\
            Kepler-359 c & $2.7^{+1.9}_{-1.4}$ & $4.1 \pm 0.5$ & $0.21^{+0.19}_{-0.11}$ & $0.2744 \pm 0.0033$ & $8.1 \pm 1.3$ & $6.3^{+4.2}_{-4.0}$ & $1.7^{+7.1}_{-4.9}$ \\
            Kepler-359 d & $2.7^{+1.8}_{-1.3}$ & $4.6 \pm 0.9$ & $0.15^{+0.19}_{-0.09}$ & $0.3329 \pm 0.0040$ & $5.5 \pm 0.9$ & $6.3^{+4.2}_{-4.0}$ & $1.4^{+7.3}_{-4.7}$ \\
            \hline
        	\end{tabular*}
        	\begin{tablenotes}
             \small
			 \item {\bf Notes.} 
			 \item \textsuperscript{a}{Mass estimates are obtained by using mass ratio posteriors from \cite{Hadden2017} unless otherwise specified and stellar mass estimates from \cite{Fulton2018}.}
			 \item \textsuperscript{b}{Values from \cite{Fulton2018}. The only parameter for which asymmetric error bars are important is stellar age.}
			 \item \textsuperscript{c}{Mass ratio posteriors, planetary radii, and stellar age from \cite{Libby-Roberts2020}.}
			 \item \textsuperscript{d}{Mass ratio posteriors and radius estimate from this study.}
			\end{tablenotes}
	    \end{center}
	\end{threeparttable}
\end{table*}

\begin{figure*}
	\centering
    \subfigure[Photospheric pressure = 10 mbar]{\includegraphics[width=0.48\linewidth]{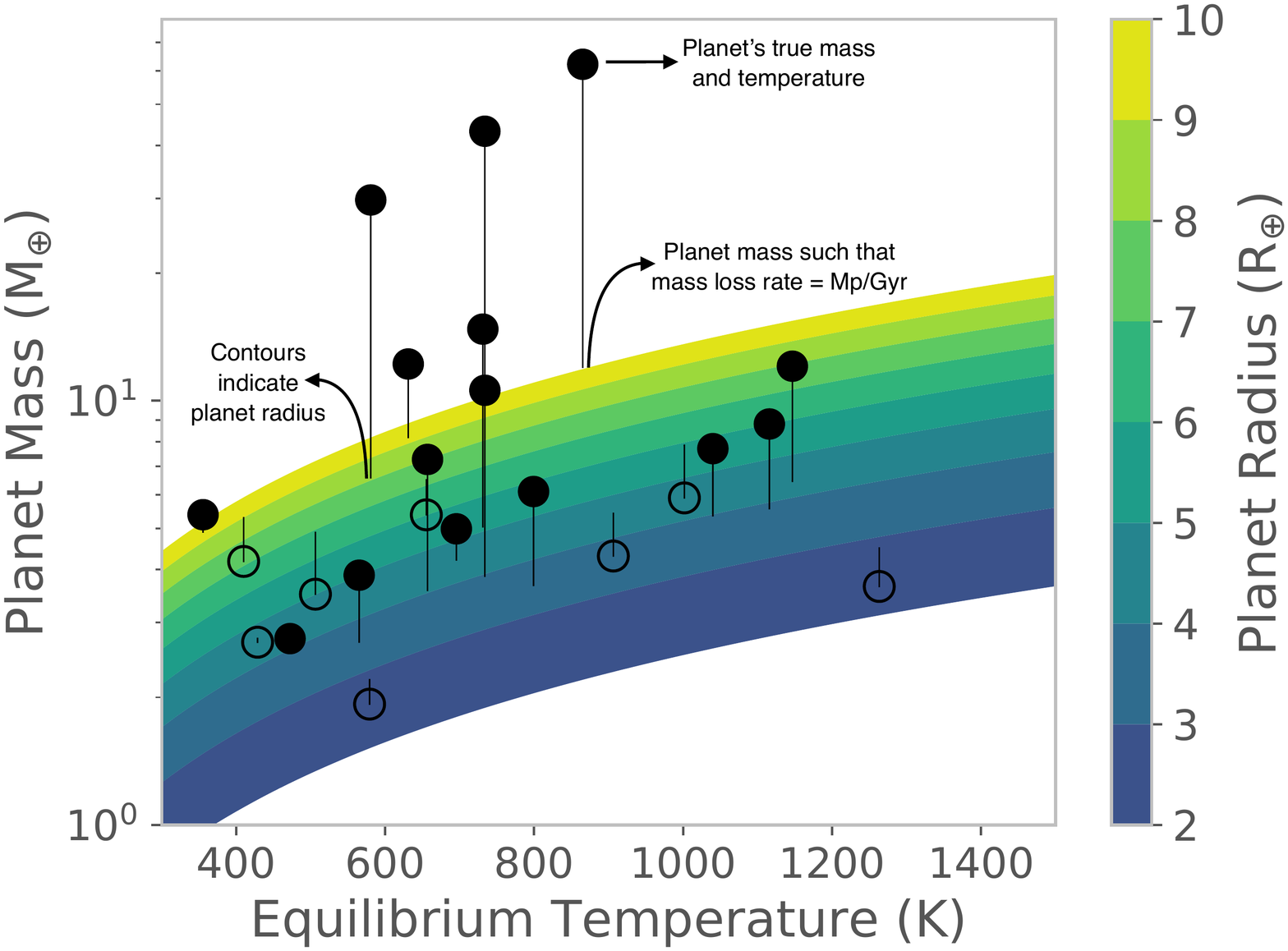}}
	\subfigure[Photospheric pressure = 10 $\mu$bar]{\includegraphics[width=0.48\linewidth]{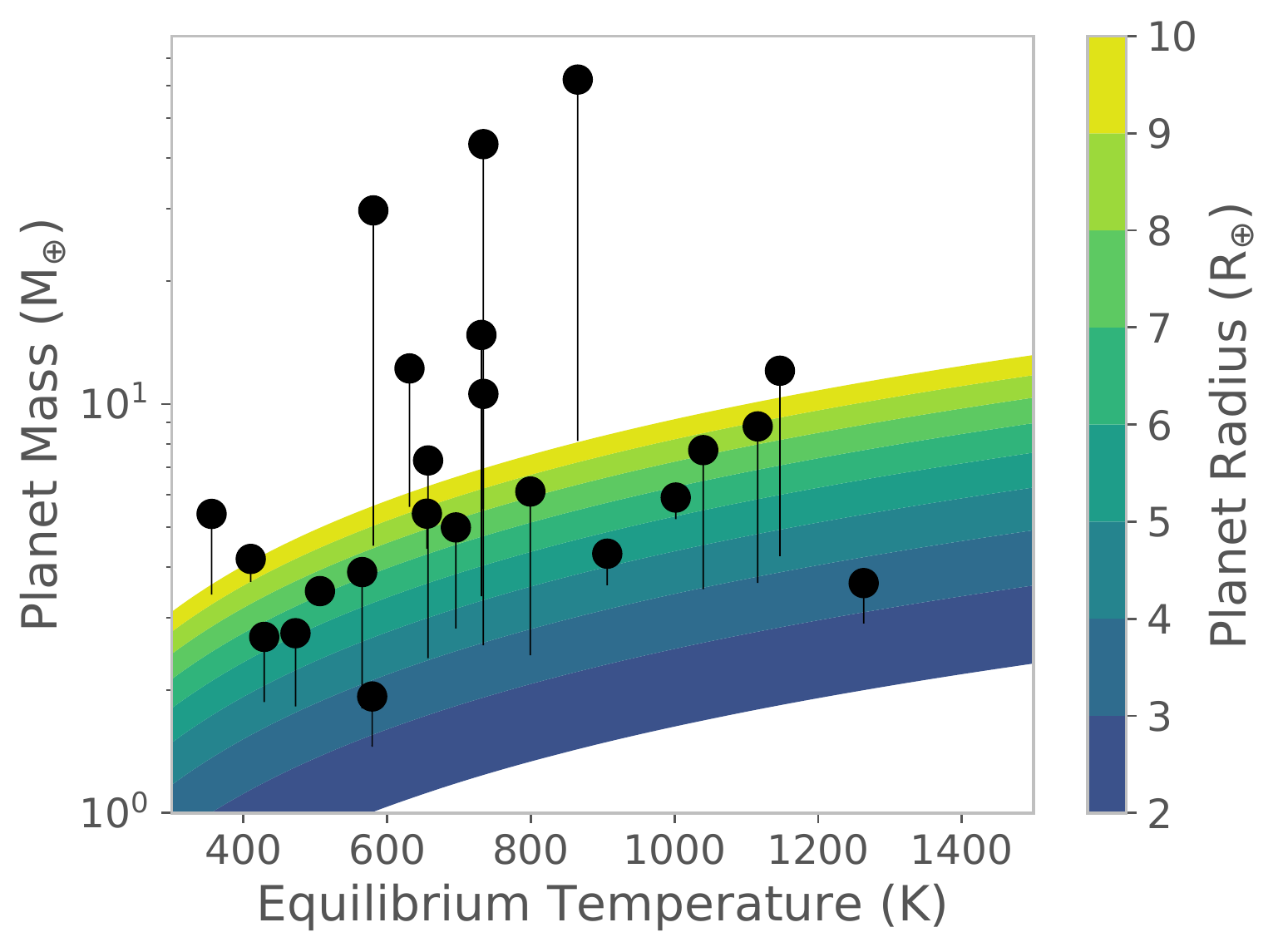}}
	\caption{Planet mass vs equilibrium temperature. The contours correspond to planetary radii for which Parker wind mass loss rates would be equal to $\mathrm{M}_{\mathrm{p}}$ Gyr$^{-1}$. Planets with a mass loss rate smaller (larger) than this value are plotted with filled (empty) circles. Vertical lines originating from the scatter points terminate at the planet mass value for which the mass loss rate would be $M_{\mathrm{p}}$ Gyr$^{-1}$ (keeping the planet radius and equilibrium temperature constant). Kepler-79d is plotted at 630 K and 5.3 $M_{\oplus}$ and changes from being an open to a filled circle when the photospheric pressure is changed.}
	\label{fig:10mbar_vs_10ubar}
\end{figure*}

We calculate the expected mass loss rate, $ \dot{M}$, for each of these planets using the isothermal Parker wind model \citep{Parker1958}\footnote{We find that the photoevaporation mass loss rate is smaller than the Parker wind mass loss rate for part of our sample that is vulnerable to catastrophic mass loss. The handful of planets for which photoevaporation dominates are quite massive (10s of M$_{\oplus}$) and the mass loss rate is too small to be significant.}:
\begin{equation}
    \dot{M} = 4 \pi r_s^2 c_s \rho_p \; \mathrm{exp}(3/2 - 2 r_s/ R_p)
    \label{eq:m_loss}
\end{equation}
where $c_s = \sqrt{k_B T_{\mathrm{eq}} / \mu}$ is the isothermal sound speed, $k_B$ is the Boltzmann constant, and $\mu$ is the mean molecular weight of the atmosphere (fixed to 2.2 here) respectively. $\rho_p$ is the atmospheric density at the measured planet radius $R_p$, and we calculate it by assuming that $R_p$ corresponds to a pressure of 10 mbar or 10 $\mu$bar (Figure~\ref{fig:10mbar_vs_10ubar}). The usage of $T_{\mathrm{eq}}$ as the local temperature here is acceptable for this pressure range. The sonic radius $r_s = G M_p / 2 c_s^2$, where $M_p$ is the planet mass and $G$ is the gravitational constant. For a range of planet masses and equilibrium temperatures, we calculate the planet radii that would yield a mass loss rate of $M_{\mathrm{p}}$ Gyr$^{-1}$. Although a planet's envelope mass is more relevant than the total mass for mass loss estimates, the envelope mass should roughly scale with the total mass\footnote{We performed the same analysis for a mass loss rate of $0.1 M_{\mathrm{p}}$ Gyr$^{-1}$ ($M_{\mathrm{env}} \sim 0.1 M_{\mathrm{p}}$ is representative of puffy planets, e.g. \citealp{Lopez2014}) and found no qualitative differences in our inferences.}. Planets in our sample are overplotted with filled ($\dot{M} < M_{\mathrm{p}}$ Gyr$^{-1}$) and empty ($\dot{M} > M_{\mathrm{p}}$ Gyr$^{-1}$) circles. For each planet's radius and equilibrium temperature, we calculate the mass, $M_{\mathrm{perpetual}}$, it would need to possess for $\dot{M}$ to be equal to $M_{\mathrm{p}}$ Gyr$^{-1}$. The values of $M_{\mathrm{perpetual}}$ are the end points of vertical lines connected to the filled and empty circles. The length of these lines is an indicator of how much larger or smaller the planets' mass loss rates are relative to $M_{\mathrm{p}}$ Gyr$^{-1}$ and therefore becomes a proxy for the atmospheric lifetimes. Figure~\ref{fig:10mbar_vs_10ubar} shows that moving the transit radius to lower pressures can significantly increase the inferred atmospheric lifetimes, resolving the apparent tension between the predicted mass loss rates and reported ages for these objects.

To estimate planet specific impact of mass loss, we calculate the atmospheric lifetime for each planet. First, we calculate the expected envelope mass fraction from the tables of \cite{Lopez2014} for a given planet mass, radius, stellar insolation, and age. Instead of assuming that the photosphere lies at 20 mbar as \cite{Lopez2014} do, we set the measured radius to correspond to a pressure of 10 $\mu$bar and calculate the 20 mbar radius by using the planetary isothermal scale height. This leads to a significantly lower envelope mass fraction estimate as our 20 mbar radius is smaller than the observed planetary radius. \cite{Lopez2014} only provide tables for envelope mass fractions between 0.01\% and 20\%. For planets that require an envelope mass fraction higher than 20\% or lower than 0.01\% to match their mass and radius, we set it to these bounding limits instead. In addition, for any planet mass or age value that lies beyond the grid limits in \cite{Lopez2014}, we set it to nearest value that is present in their tables.

The mass loss rate for each planet is calculated using both Parker wind and photoevaporation models and we then use the dominant mechanism to estimate the atmospheric lifetime. Mass loss due to photoevaporation is calculated using the standard energy limited prescription \citep[e.g.][]{Watson1981, Salz2015}:
\begin{equation}
    \dot{M} = \frac{\epsilon \pi F_{\mathrm{XUV}} R_{\mathrm{XUV}}^3}{G M}
\end{equation}
where $F_{\mathrm{XUV}}$ is the high energy flux received from the star (taken from \citealp{Ribas2005}\footnote{Although the relationship provided by \cite{Ribas2005} is applicable only to G stars, in the absence of such information for other stellar types, we use it for all the stars in our sample. The fact that we use the planet's semimajor axis to estimate $F_{\mathrm{XUV}}$ does compensate for differences in spectra of different stellar types to a certain extent. For example, for later type stars, the bolometric luminosity at a given semimajor axis is lower but the fraction of energy emitted in the XUV is higher compared to G stars.}), $R_{\mathrm{XUV}}$ is the typical radius at which this flux is absorbed (radius corresponding to pressure of 10 nbar, estimated using the planet's isothermal scale height), $G$ is the gravitational constant, $M$ is the planet's mass, and $\epsilon = 0.1$ is an `efficiency' parameter that encapsulates the complicated process of conversion of photon energy to kinetic and thermal energy of the wind. We obtain posteriors for the atmospheric lifetime by using posterior distributions for all the relevant input parameters (shown in Table~\ref{table:planet_properties}) except the atmospheric mass fraction. Calculating the atmospheric mass fraction for all our posterior samples would impose a large computational overhead. Using a smaller sample size, we find that for the majority of planets in our sample, the 84th percentile and the 16th percentile value for the atmospheric mass fraction differ only by a factor of 2. This factor is more than an order of magnitude only for the planets for which we cannot place tight meaningful constraints on the atmospheric lifetime due to imprecise planetary properties.

\begin{figure}
    \centering
    \includegraphics[width=\linewidth]{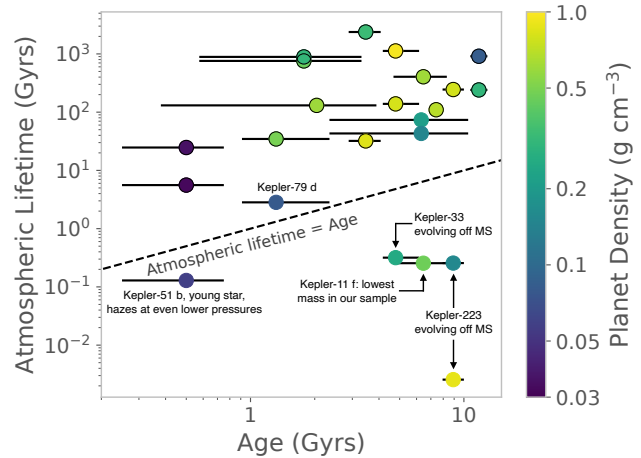}
    \caption{Atmospheric lifetime versus planet (stellar) age obtained from \cite{Fulton2018}. Atmospheric lifetime is calculated by dividing the envelope mass fraction inferred from \cite{Lopez2014} with the mass loss rate from either photoevaporation or Parker wind (whichever one is larger, photoevaporation dominated planets circled with black outline). Due to large uncertainties in planetary properties, the uncertainties on the atmospheric lifetime (not shown) are large enough that most planets lie $< 1 \sigma$ away from the atmospheric lifetime = age dashed line.}
    \label{fig:atm_lifetime_age}
\end{figure}

Figure~\ref{fig:atm_lifetime_age} shows how the median atmospheric lifetime of our sample compares with the planetary (stellar) ages. A majority of the planets that have atmospheric lifetimes longer than their ages are most susceptible to photoevaporative mass loss. In contrast, Parker winds drive the envelope loss for all the low density planets that have atmospheric lifetime shorter than their ages. It is worth noting that the uncertainty on the atmospheric lifetime (not shown for clarity) for most planets is so large that they are $< 1 \sigma$ away from the atmospheric lifetime and age equality. We therefore caution the reader to not use this plot to draw inferences regarding any trends in atmospheric lifetime with host star age or planet density. This plot simply shows that the median atmospheric lifetimes for all but a handful of these planets are larger than their present-day ages. The calculated atmospheric lifetimes and the corresponding $1 \sigma$ uncertainties are listed in Table~\ref{table:planet_properties}.

Within this parameter space, Kepler-51 b, Kepler-223 b and d, Kepler-33 d, and Kepler-11 f stand out as some of the shortest-lived puffy planets. For Kepler-51 b, we find that using the updated mass values from \cite{Libby-Roberts2020} increases the median atmospheric lifetime by almost 3 orders of magnitude for just a 50\% increase in median planet mass. This is a result of the exponential sensitivity of Parker wind mass loss rate to planet mass. Similarly, Kepler-11 f's low atmospheric lifetime is a consequence of the fact that it is the lowest mass puffy planet in our sample. The continued cooling and contraction of Kepler-51 b (which likely still possesses its heat of formation) as it ages would also reduce the inferred mass loss rate at later times \citep[see][]{Libby-Roberts2020}. Interestingly, Kepler-223 and Kepler-33 have inflated stellar radii and are evolving off the main sequence (in fact it is their evolution off the main sequence that allows us to measure their ages much more precisely than for typical main sequence stars). Their planets therefore may have recently started losing significant mass in response to the increasing luminosity of their host stars. The planets' current mass loss rates are likely to be significantly larger than during the main sequence phase of the host stars. Hence, the incompatibility between their atmospheric lifetimes and age does not necessarily pose a contradiction. For planets with very high mass loss rates (such as Kepler-51 b), the outflowing wind may loft hazes to pressures as low as $10-100$ nbar \citep{Gao2020}, increasing their apparent size and the corresponding atmospheric lifetime by $2-3$ orders of magnitude relative to the $10 \; \mu$bar case.

Another physical explanation that has recently been used to explain the large radii of super-puffs without invoking large envelope mass fraction is tidal heating \citep{Millholland2019}. Here, we briefly comment on the consequences of this model for the atmospheric lifetime of super-puffs, with a particular emphasis on the case of Kepler-79d. Supplementing the internal luminosity of the planet with tidal heating leads to a significant decrease in envelope mass fraction (e.g., a factor of 3 decrease for Kepler-79d for a haze free atmosphere; \citealp{Millholland2019}) without necessarily affecting the atmospheric mass loss rate significantly. Atmospheres of tidally-heated super-puffs that are haze-free or have a haze at $10$ mbar are then even more vulnerable to envelope loss than super-puffs without tidal heating. For an atmospheric haze at 10 $\mu$bar, our CARMA models indicate that the envelope mass fraction for a tidally heated Kepler-79d ($T_{\mathrm{int}} = 300$ K) is nearly 30 times lower than that for the model without tidal heating ($T_{\mathrm{int}} = 75$ K). This order of magnitude decrease in the envelope mass fraction almost nullifies the advantage of placing hazes at $10 \mu$bar to reconcile mass loss history with atmospheric lifetime. This suggests that tidal heating may not play a key role in the super-puff story, especially for longer period super-puffs such as Kepler-79d. Instead, tidal heating is likely to be more important for the population of low-density planets with close-in orbits (e.g. $P \lesssim 30$ days) and moderate eccentricities, as shown in \cite{Millholland2020}. It is worth noting that our analysis only provides a weak and indirect constraint on models with tidal heating. Nevertheless, it could be a valuable and independent semi-observational test for such models in the future.

It has previously been suggested that the high inferred mass loss rates for super-puffs might stem from incorrect mass loss rate prescriptions or incorrect values for planet mass or temperature \citep{Cubillos2017}. We note two caveats in the mass loss prescriptions we use in this work. Firstly, we assume an isothermal Parker wind structure for the outflow driven by the bolometric flux of the star and this gives us an upper limit on the mass loss rate. If the outflow cools as it propagates outwards, the mass loss rate would be lower. Secondly, the exact mass loss rate is sensitive to the assumed atmospheric structure. Mass loss models often assume an inner adiabat with adiabatic index of 7/5 that either extends to the photosphere \citep{Gao2020} or transitions to an isothermal layer or an wind-launching surface \citep{Wang2018}. Such structure maximizes the rate of mass loss as the atmospheric mass is outwardly concentrated \citep{Lee2018}. At formation, a more realistic adiabatic index for super-puffs is $\sim 1.2$ which gives rise to inwardly concentrated mass profile \citep{Lee2014, Lee2016}. Self-consistent treatment with the evolving inner adiabat for super-puffs is yet to be conducted. Notwithstanding these limitations, we note that atmospheric mass loss models have been relatively successful to date in matching the measured mass loss rates for transiting gas giant planets \citep[e.g.][]{Salz2016, Salz2018, Odert2019} as well as the bimodality in the radius distribution of the sub-Neptune-sized planets in the \emph{Kepler} sample \citep{Lopez2013, Fulton2017, Owen2017}, making it unlikely that our mass loss rates are incorrect by many orders of magnitude. 

Similarly, it would require an implausibly large reduction in the planetary temperature (corresponding to albedos as high as those of the icy moons in the solar system; see also \citealp{Cubillos2017}) to reconcile the predicted mass loss rates with planetary ages. Mass estimates, on the other hand, can be quite uncertain for many of these planets (the median fractional uncertainty in planet mass for our sample is 21\%) and the mass loss rate is quite sensitive to the assumed value (Equation~\ref{eq:m_loss}). It is evident from Table~\ref{table:planet_properties} that planets with uncertain masses have a $1 \sigma$ range in atmospheric lifetime that spans many orders of magnitude. Therefore, until the masses of these planets are more precisely determined\footnote{With the advent of the \emph{Gaia} era, the planet mass uncertainties are dominated by the uncertainties in the mass ratios obtained from TTV fits rather than the uncertainties in stellar masses. The only exceptions to this in our sample are Kepler-9 (fractional stellar mass uncertainty twice that of mass ratio's) and Kepler-36 (fractional stellar mass and mass ratio uncertainties comparable).}, we cannot dismiss outright the possibility that uncertainties in planet mass might also contribute to the apparently short atmospheric lifetimes of some super-puffs.

\section{Future Directions and Conclusions}
\label{sec:future_conclusions}
In this work we present new observational constraints on the properties of Kepler-79d, a quintessential super-puff. Our revised planet mass further cements its status as a low density planet. The availability of \emph{Kepler and HST} data as well as the relatively low activity level of the host star allows us to constrain the atmospheric scattering signature across optical and infrared wavelengths. We find that the transmission spectrum does not contain any statistically significant absorption signatures and is consistent with a flat line. Our data rule out both metal-enriched cloud free models and metal-poor models with only collision-induced absorption and Rayleigh scattering from hydrogen and helium. We therefore conclude that Kepler-79d most likely hosts a high-altitude haze. We use CARMA models that incorporate haze entrainment by an outflowing  wind to show that the resultant transmission spectrum provides a reasonable fit to the data. The shift of the slant photosphere to lower pressures reduces the amount of primordial gas required to match the planet's bulk density and lengthens the atmospheric lifetime by decreasing the mass loss rate. We also show that this effect of reducing the photospheric pressure significantly affects our mass loss inferences for low density planets in the \emph{Kepler} sample and tends to reconcile planetary ages with their current primordial envelope content. 

Super-puffs are an enigmatic and exciting sub-population of the \emph{Kepler} planets. Their minuscule bulk densities pose difficult challenges for planet formation theories and atmospheric evolution models. It remains to be seen whether they are ringed planets \citep{Piro2020}, typical super-Earths with large internal heat fluxes \citep{Millholland2019}, or planets with large primordial atmospheric content \citep[e.g.][]{Lopez2014, Lee2016}. Hazes seem likely to be a part of the super-puff story, as they may simultaneously explain their transmission spectra \citep{Libby-Roberts2020} and the mass loss history \citep{Wang2019, Gao2020}. High-altitude hazes also have the advantage of offering a more universal explanation for the properties of super-puff atmospheres compared to the other hypotheses. Follow up studies are critical for making further inferences but the dimness of the \emph{Kepler} super-puff hosts has hitherto limited our ability to do so. This is likely to change with the advent of the \emph{JWST} era with its increased sensitivity and access to a broad and optimal wavelength range for targeting super-puff atmospheres. \emph{JWST} observations will provide data that are truly diagnostic and powerful at distinguishing between different models. The discovery of super-puffs around bright nearby host stars will also provide us with significantly more favorable targets for atmospheric characterisation \citep[e.g.][]{Santerne2019}.

\section*{Acknowledgements}
We are indebted to the reviewer for their apt and useful suggestions that improved this manuscript. This work is based on observations from the Hubble Space Telescope, operated by AURA, Inc. on behalf of NASA/ESA. Support for this work was provided by NASA through Space Telescope Science Institute grants GO-14260 and GO-15138. P. Gao and I. Wong acknowledge the generous support of the Heising-Simons Foundation via the 51 Pegasi b fellowship in Planetary Astronomy. E.B.F. acknowledges support from the Penn State Eberly College of Science and Department of Astronomy \& Astrophysics and the Center for Exoplanets and Habitable Worlds. We would also like to thank Jim Fuller, Sarah Millholland, Dave Stevenson, and Shreyas Vissapragada for helpful discussions and feedback.

\software{\texttt{Astropy} \citep{astropy:2013, astropy:2018}, \texttt{BATMAN} \citep{Kreidberg2015}, \texttt{emcee} \citep{Foreman-Mackey2012}, \texttt{LDTk} \citep{Parviainen2015}, \texttt{Matplotlib} \citep{Hunter:2007}, \texttt{NumPy} \citep{oliphant2006guide, van2011numpy}, \texttt{PLATON} \citep{Zhang2019}}

\newpage
\bibliography{manuscript}
\bibliographystyle{apj}

\end{document}